\documentclass[aps,prx,twocolumn,superscriptaddress,nofootinbib,longbibliography]{revtex4-1}
\usepackage{graphicx}
\usepackage{amssymb}
\usepackage{amsmath}
\usepackage{amssymb}
\usepackage{amsfonts}
\usepackage{mathtools}
\usepackage{bm}
\usepackage{braket}
\usepackage[dvipsnames]{xcolor}
\usepackage[colorlinks=True,citecolor=blue,urlcolor=blue,linkcolor=red]{hyperref}
\usepackage{hypcap}
\newcommand{\up}{\uparrow}
\newcommand{\dn}{\downarrow}

\newcommand{\nh}{N_\mathcal{H}}
\newcommand{\E}{\mathcal{E}}
\renewcommand{\P}{\mathcal{P}}

\newcommand{\typ}{\mathrm{typ}}
\newcommand{\comment}[1]{}

\newcommand{\pd}{\phantom\dagger}
\newcommand{\w}{\omega}

\newcommand{\ssh}[2][I]{s^{\scriptscriptstyle(#1)}_{#2}}
\newcommand{\tot}{\mathrm{tot}}

\begin{document}

\author{Sthitadhi Roy}
\email{sthitadhi.roy@chem.ox.ac.uk}
\affiliation{Department of Chemistry, Physical and Theoretical Chemistry, Oxford University, South Parks Road, Oxford, OX1 3QZ, United Kingdom}
\affiliation{ Rudolf Peierls Centre for Theoretical Physics, Clarendon Laboratory,
Oxford University, Parks Road, Oxford OX1 3PU, United Kingdom}

\author{David E. Logan}
\email{david.logan@chem.ox.ac.uk}
\affiliation{Department of Chemistry, Physical and Theoretical Chemistry, Oxford University, South Parks Road, Oxford, OX1 3QZ, United Kingdom}
\affiliation{Department of Physics, Indian Institute of Science,
Bangalore 560 012, India}

\date{\today}

 \title{Fock-space correlations and the origins of many-body localisation}
\begin{abstract}
We consider the problem of many-body localisation on Fock space, focussing on the essential features of the  Hamiltonian which stabilise a localised phase. Any many-body Hamiltonian has a canonical representation as a disordered tight-binding model on the Fock-space graph. The underlying physics is however fundamentally different from that of conventional Anderson localisation on high-dimensional graphs because the Fock-space graph possesses non-trivial correlations. These correlations are shown to lie at the heart of whether or not a stable many-body localised phase can be sustained in the thermodynamic limit, and a theory is presented for the conditions the correlations must satisfy for a localised phase to be stable. Our analysis is rooted in a probabilistic, self-consistent mean-field theory for the local Fock-space propagator and its associated self-energy; in which the Fock-space correlations, together with the extensive nature of the connectivity of Fock-space nodes,  are key ingredients. The origins of many-body localisation in typical local Hamiltonians where the correlations are strong, as well as its absence in uncorrelated random energy-like models,  emerge as predictions from the same overarching theory. To test these, we consider three specific microscopic models, first establishing in each case the nature of the associated Fock-space correlations. Numerical exact diagonalisation is then used to corroborate the theoretical predictions for the occurrence or otherwise of a stable many-body localised phase; with mutual agreement found in each case.
\end{abstract}

\maketitle

\tableofcontents

\section{Introduction}

\begin{figure}
\includegraphics[width=\columnwidth]{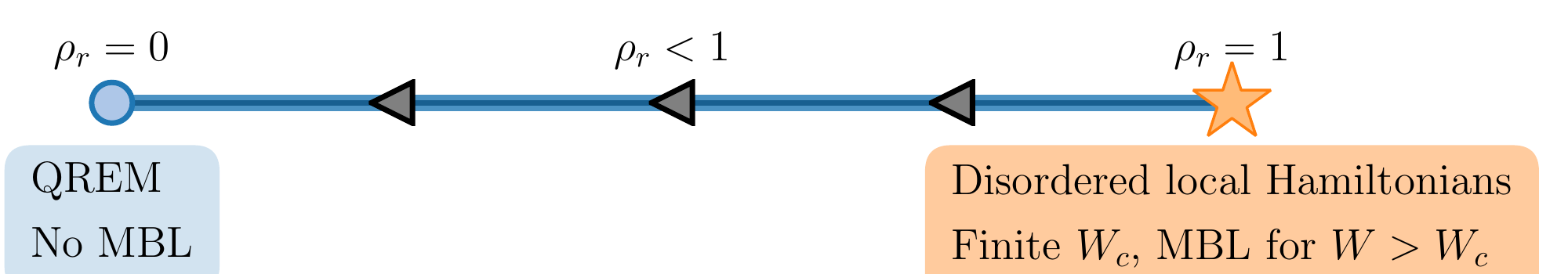}
\caption{Schematic of primary result of the paper. 
Disordered local models where the Fock-space site energies are maximally correlated, such that their rescaled covariance 
$\rho_{r}$ as a function of Hamming distance $r$ satisfies $\rho_r=1$ for any finite $r$, host a many-body localised phase. By contrast, models with $\rho_r=0$ where correlations are absent, such as the quantum random energy model (QREM), do not.  Any intermediate case, $\rho_r<1$, falls into the latter class and does not host a localised phase. }
\label{fig:schematic-pd}
\end{figure}

The problem of localisation in interacting quantum many-body systems has lately been a central theme in condensed matter and statistical physics, under the banner of many-body localisation~\cite{basko2006metal,gornyi2005interacting,oganesyan2007localisation,znidaric2008many,pal2010many} (for reviews see Refs.~\cite{nandkishore2015many,alet2018many,abanin2019colloquium} and references therein). In addition to the important question of the fate of single-particle Anderson localisation~\cite{anderson1958absence} upon the introduction of interactions, many-body localised systems are of fundamental interest due to their violation of the eigenstate thermalisation hypothesis (ETH). In generic ergodic systems, ETH explains the emergence of equilibrium thermodynamics and statistical mechanics, since eigenstate expectation values of local observables depend only on a few macroscopic state variables such as eigenenergies~\cite{deutsch1991quantum,srednicki1994chaos,deutsch2018eigenstate}.

Many-body localised systems by contrast are specified by an extensive set of (quasi-)local integrals of motion~\cite{serbyn2013local,huse2014phenomenology,ros2015integrals,rademaker2016explicit,imbrie2017local}, and hence 
fall outside the paradigm of conventional statistical mechanics. This was also understood as the fundamental mechanism 
underlying the absence of transport, and logarithmic growth of entanglement, characteristics of many-body localised 
phases~\cite{znidaric2008many,bardarson2012unbounded,serbyn2014quantum}. Much of the phenomenology of such systems, as well as theoretical studies of the nature of the disorder-induced localisation transition, 
has naturally centred on local models and their behaviour in real-space. These include both analytical and extensive numerical studies of disordered spin- and fermionic-chains, as well phenomenological renormalisation group studies~\cite{pal2010many,lev2015absence,luitz2015many,imbrie2016many,kjall2014many,vosk2015theory,potter2015universal,zhang2016many,dumitrescu2017scaling,thiery2018many-body,goremykina2019analytically,dumitrescu2018kosterlitz,morningstar2019renormalization}.

Nevertheless, since the problem is fundamentally a many-body one, the underlying Fock space offers a seemingly natural framework within which to study and understand it~\cite{logan1990quantum,altshuler1997quasiparticle,MonthusGarel2010PRB,logan2019many,roy2019self,pietracaprina2016forward,mace2019multifractal,pietracaprina2019hilbert,roy2018exact,roy2018percolation,ghosh2019manybody,detomasi2019dynamics}. More recently, it has been shown that this is indeed the case.
Analytical approaches to many-body localisation, based on local~\cite{logan2019many,roy2019self} and non-local 
Fock-space propagators~\cite{pietracaprina2016forward}, have been developed. Numerical studies have also shed light on the the Fock-space structure of many-body localised eigenstates~\cite{mace2019multifractal,pietracaprina2019hilbert}. Investigating the problem on the Fock space also instigated the discovery of an exactly soluble classical percolation proxy for the many-body localisation transition~\cite{roy2018exact,roy2018percolation}, which gave a range of new insights, including  possible critical exponents for the transition.

The general Fock-space properties of Hamiltonians hosting many-body localisation thus merit understanding. 
Universal properties of ergodic 
phases are generally well understood within the framework of random matrix theory~\cite{deutsch1991quantum,srednicki1994chaos,deutsch2018eigenstate,mehta2004random}.
It is equally natural to enquire about the general conditions under which a random many-body Hamiltonian can host a stable localised phase.

An inevitable characteristic of many-body Hamiltonian matrices is the presence of correlations between the matrix elements. In this work we focus on correlations in the diagonal elements of the Hamiltonian,  and ask what form the correlations must satisfy for a stable localised phase to be sustained. We obtain a precise criterion on the correlations: for two states separated by a finite (Hamming) distance on the Fock space, the difference between the corresponding diagonal elements, when rescaled with the width of the density of states, must vanish in the thermodynamic limit. This constitutes the primary message of the paper, and is summarised in Fig.~\ref{fig:schematic-pd} (itself discussed in greater detail below).

An overview of the paper and its substantive points is first given,  Sec.\ \ref{sec:overview}. 
General characteristics of the Fock-space graphs considered are discussed in Sec.~\ref{sec:FSgraph}, before
we turn in Sec.~\ref{sec:fockspacelocalisation} to a self-consistent mean-field theory for localisation on the Fock space, and the criterion for many-body localisation which arises from it. The analytical predictions are then tested and confirmed in Sec.~\ref{sec:microscopic} by comparison to numerical results for three different microscopic models, obtained by exact diagonalisation and employing well-established diagnostics for identifying ergodic and many-body localised phases. We conclude with a brief summary and discussion, Sec.~\ref{sec:discussion}.

\section{Overview}
\label{sec:overview}

This paper centres on a simply-stated question: what are the minimal properties required for a disordered many-body system to sustain a stable localised phase? Since any Hamiltonian can be represented as a tight-binding model on an associated Fock-space graph (or `lattice'), this question is naturally and conveniently addressed on that graph.
The Hamiltonian for a disordered many-body system with $N$ degrees of freedom is typically characterised by 
a polynomially large (in $N$) number of independent microscopic energy scales, which could be coupling constants, local fields etc. On the other hand, the dimension of the Fock-space graph, $\nh$, is exponentially large in $N$.
This dichotomy in scale inevitably means that matrix elements of the many-body Hamiltonian are mutually correlated. 
One  main aim of this work is to identify a criterion that these correlations must satisfy in order that a many-body localised phase be stable.

To set up the theory on  general grounds, we begin in Sec.~\ref{sec:FSgraph} by  identifying some important
characteristics of Fock-space graphs corresponding to generic  many-body systems. Any graph has two essential constituents, nodes and links. The nodes, which we refer to as Fock-space sites and denote by $I$, correspond to the many-body basis states; each of  which has an associated on-site energy $\E_I$, corresponding simply to the diagonal elements of the many-body Hamiltonian. Links between these nodes correspond by contrast to off-diagonal matrix elements of the Hamiltonian, and represent  hopping between the Fock-space sites in the effective tight-binding problem.
The matrix elements corresponding to a link are $\mathcal{O}(1)$ numbers, and the average connectivity of a 
Fock-space site is perforce extensive in the system size $N$. For simplicity and clarity, we consider 
the case where every Fock-space site has a connectivity of  precisely $N$, and all non-vanishing off-diagonal matrix elements have equal magnitude (though these conditions are readily relaxed). The links also confer the graph with a distance metric; the Hamming distance between two nodes on the graph, defined  as the shortest path between them following the links.

The correlations in $\{\E_I\}$ are  a crucial feature of the Fock-space graphs. Over an ensemble of disorder realisations, 
the entire family of disordered Hamiltonians is fully specified by the $\nh$-dimensional joint distribution of 
$\{\E_I\}$, which encodes these correlations. We argue that this distribution, $\P^d_{\nh}$, is generically an $\nh$-dimensional (multivariate) Gaussian,  the covariance matrix of which, $\mathbf{C}_d$, satisfies two important properties:  its elements (i) scale as $N$, and (ii) depend only on the Hamming distance, $r$, between the two Fock-space sites in question.

The distribution $\P^d_{\nh}$ is over disorder realisations. For any given disorder realisation, one can also consider a 
$\mathcal{D}$-dimensional distribution  over the Fock-space sites, $\P_\mathcal{D}^F$, provided $\mathcal{D}\ll\nh$;
with a finite covariance for this distribution reflecting  the intrinsic configurational disorder in the problem.
We argue in particular that for models with local and short-ranged interactions, $\P_\mathcal{D}^F$ is also a 
multivariate Gaussian distribution, with its covariance matrix again depending only on the Hamming distance.
In consequence, there exists a joint distribution $\P_\mathcal{D}$ for the Fock-space site energies over both disorder realisations and the Fock space, which is a convolution  of the two distributions, 
$\P_\mathcal{D} = \P^d_\mathcal{D} \ast \P_\mathcal{D}^F$.

In Sec.~\ref{sec:fockspacelocalisation} we turn to a self-consistent theory of localisation on the Fock space.
A probabilistic mean-field theory is set up for the imaginary part of the local self-energy on the Fock space,
$\Delta_I(\omega)$~\cite{logan2019many,roy2019self}. Physically, $\Delta_I(\omega)$ is proportional to the rate of loss of probability from Fock-space site $I$ into  states of energy $\w$. For a delocalised phase in the thermodynamic limit,
$\Delta_I/\sqrt{N}$  is finite with unit probability over disorder realisations, while it vanishes with unit probability in a localised phase, and hence acts as diagnostic of the transition. To enable the thermodynamic limit to be taken, all many-body energies  must in fact be rescaled by  $\sqrt{N}$~\cite{logan2019many,roy2019self}.
The mean-field theory is constructed using a Feenberg renormalised perturbation series~\cite{feenberg1948note,Economoubook},  with the distribution 
$\P_\mathcal{D}$ and the extensive connectivity of Fock space as core ingredients;
and our focus in practice being band centre states in the many-body spectrum (density of states).
Within this framework, the distribution of the local self-energy can be obtained self-consistently; 
with the breakdown of the underlying self-consistency equation signalling the breakdown of the
phase and an ensuing transition. As all many-body energies are rescaled by $\sqrt{N}$, it is the rescaled covariance, 
$C(r)/N$, which admits a well-defined thermodynamic limit and encodes the criterion for a localised phase to be stable. Since $C(0)\propto N$, the theory is thus  formulated in terms of the reduced covariance, 
\begin{equation}
\nonumber
\rho_{r}^{\pd} ~=~C(r)/C(0) .
\end{equation}

We find that \emph{for a stable many-body localised phase to persist, it is necessary that $\rho_r\to 1$ for finite
Hamming distances $r$, in the thermodynamic limit $N\to\infty$.} 
This is  our most central result (see Fig.~\ref{fig:schematic-pd}).
We elucidate it by considering first the two extreme limits of correlations. In Sec.~\ref{subsec:maximally} we discuss 
what we term the \emph{maximally correlated limit}, where $\rho_r\to 1$ for all finite $r$.
This case arises typically in models with local interactions, because Fock-space sites separated by a finite Hamming distance also have their Fock-space site energies differing by a finite $\mathcal{O}(1)$ energy.
Hence, on rescaling by $\sqrt{N}$, these differences vanish in the thermodynamic limit, and the $\E_I/\sqrt{N}$ for such 
sites are completely slaved to each other. In this case we find that a self-consistent 
localised solution is indeed possible at finite values of disorder strength, whence a localised phase is stable.

In Sec.~\ref{subsubsec:QREMlimit} we turn to the opposite limit, which we  refer to as the \emph{random energy limit}~\cite{derrida1980random}.
Here the Fock-space site energies are completely uncorrelated, with $\rho_r=0$ for all non-zero $r$.
In this case we find that a localised phase is unstable for any non-zero hopping between the Fock-space sites.
Note that our theoretical predictions in these two limits accord with earlier studies of many-body localisation in short-ranged models~(see \cite{nandkishore2015many} for review and further references), and the quantum random energy model (QREM)~\cite{laumann2014many,baldwin2016manybody}. Finally, in Sec.~\ref{subsubsec:arbitcor} we treat the general case of $0<\rho_r<1$, which is found to fall in the QREM class, such that a localised phase is not possible. This leads us to conclude that a many-body localised phase is stable only in the class of models that we term maximally correlated, \textit{i.e.} $\rho_r\to 1$ for finite $r$ in the thermodynamic limit.

These theoretical predictions clearly call for comparison with results from microscopic models, in particular using
established numerical methods which are free from the approximations inevitable in a mean-field theory.
Section~\ref{sec:microscopic} is devoted to such an analysis, with three distinct models considered: a disordered 
nearest-neighbour quantum Ising chain widely used as a model for many-body localisation (Sec.~\ref{subsection:NNTFI}), disordered $p$-spin models with finite $p$ (Sec.~\ref{subsection:pspinmods}), and a model we name the ExpREM, which is a QREM modified to have  the covariance of its Fock-space site energies decay exponentially with Hamming distance 
(Sec.~\ref{subsection:expREM}).

The analysis of Sec.\ \ref{sec:microscopic} consists first of showing that the distributions $\P_\mathcal{D}(\{\E_I\})$ 
are indeed $\mathcal{D}$-dimensional Gaussians,  with a covariance matrix whose elements depend only on the Hamming 
distance. This is demonstrated numerically for the disordered Ising chain, while for the $p$-spin models we show
it analytically. Analytical results for the covariance matrix are also derived for both models (and agree very well with numerics). These covariance matrices are such that both models fall into the maximally correlated class 
discussed in Sec.~\ref{subsec:maximally}, and are thus expected to  possess a localised phase. This is indeed consistent with numerical results obtained from Exact Diagonalisation,  using the three complementary diagnostics of eigenvalue level-spacing ratios, participation entropies of eigenstates on the Fock space, and bipartite entanglement entropies.
The ExpREM model by contrast does not fall into the maximally correlated class; and exact diagonalisation results are consistent with the theoretical prediction that such a model does not possess a many-body localised phase in the middle of the spectrum.

We conclude the paper in Sec.~\ref{sec:discussion} by discussing some key points pertaining to localisation in 
high-dimensional graphs,  as well as the connection of the present theory to other complementary approaches to 
understanding the physics of many-body localisation on the Fock space.
For convenience, a glossary of symbols commonly used in the paper is also given in Appendix \ref{section:glossary}.


\section{Characteristics of Fock-space graphs}
\label{sec:FSgraph}

Fock space provides a natural framework for studying many-body localisation, since a generic many-body Hamiltonian maps exactly onto a tight-binding model on the Fock space, with correlated on-site disorder on its constituent sites.
We represent this Fock-space Hamiltonian as 
\begin{equation}
	H = \underbrace{\sum_{I=1}^{N_\mathcal{H}}\mathcal{E}_I\ket{I}\bra{I}}_{H_\text{diag}} + \underbrace{{\sum_{I,K}}^{\prime}T_{IK}\ket{K}\bra{I}}_{H_\text{offdiag}\equiv H_\text{1}},
	\label{eq:ham-fs}
\end{equation}
(where $^{\prime}$ denotes $K\neq I$). Here $N_\mathcal{H}$ is the Fock-space dimension, with $\{\ket{I}\}$ the set of basis states spanning the Fock space, and $\{\mathcal{E}_I\}$ the on-site energies on the Fock space, which are typically correlated in a non-trivial fashion. $\{T_{IK}\}$ is the set of Hamiltonian matrix elements connecting pairs of basis states, representing hoppings on the Fock space.
As the central theme of this paper is how the  statistics and correlations of $\{\mathcal{E}_I\}$ can interplay with
the $\{T_{IK}\}$ and lead to many-body localisation, let us now define these notions concretely.
While the above construction is general, as an archetype of quantum many-body systems we will focus for concreteness on interacting quantum spin-1/2 systems; with the spins denoted by the set of Pauli matrices $\{\sigma_{\ell}^{\mu}~|~\mu=x,y,z\}$, where $\ell=1,2,..N$ denotes the $N$ underlying real-space sites.  Without loss of generality we choose our basis states to be $\sigma^z$-product states. Note too that the Fock-space dimension  $\nh$ is perforce exponentially large in $N$.

A general and seemingly quite ubiquitous property of many-body Hamiltonians is that the \emph{average  connectivity} on the Fock space graph scales linearly with the system size (number of spins), $N$. We define the average effective connectivity on the graph as $N_\mathcal{H}^{-1}{\sum}^{\prime}_{I, K}\vert T_{IK}\vert$. 
Since $T_{IK}$ is the matrix element connecting two basis states via generically local spin-flips, it is an $\mathcal{O}(1)$ number. At the same time, for a system with $N$ spins, there are $\mathcal{O}(N)$ different local spin-flips possible from a given configuration. Hence the average connectivity of the graph is also $\mathcal{O}(N)$. 
Exceptions are indeed possible, e.g.\ power-law interacting systems wherein the scaling is sensitive to the decay exponent of the power-law interactions~\cite{roy2019self}; but for brevity we do not explore this direction in the current paper.

In the rest of the paper, as a representative model for $H_\text{1}$, we consider 
\begin{equation}
	H_\text{1} = \Gamma\sum_{\ell=1}^N\sigma^x_\ell.
	\label{eq:ham-offdiag}
\end{equation}
This is one of the simplest examples that satisfies the properties mentioned above. Its Fock-space dimension is
$\nh =2^{N}$. On the Fock space,  the neighbours of a site $\ket{I}$, under $H_\text{1}$ in Eq.~\eqref{eq:ham-offdiag}, are related to $\ket{I}$ by exactly one spin-flip. Hence the connectivity of every Fock-space site is precisely $N$, and its distribution a $\delta$-function centred on $N$. We emphasise that the model in Eq.~\eqref{eq:ham-offdiag} is not special, and only the aforementioned general properties of the connectivities are required.\footnote{Another Hamiltonian 
commonly studied in MBL is the disordered XXZ-chain, where $H_\text{1} = \Gamma\sum_{\ell=1}^N[\sigma^x_\ell\sigma^x_{\ell+1}+\sigma^y_\ell\sigma^y_{\ell+1}]$. In this case the connectivity distribution is a Gaussian with mean $\propto N$ and standard deviation $\propto \sqrt{N}$~\cite{welsh2018simple}.}

As a metric of distance on the Fock space, we use the \emph{Hamming distance} $r_{IK}$, defined as  the number of 
real-space sites on which the spin-orientation between the two spin-configurations $I$ and $K$ is different. For the links on the Fock-space graph generated by $H_\text{1}$ of Eq.~\eqref{eq:ham-offdiag}, this distance is also the shortest distance on the graph.

Let us now turn to the statistics and correlations of $\{\mathcal{E}_I\}$. While every $\mathcal{E}_I$ is random, the origin of correlations between them can be understood from the fact that although there are exponentially large (in $N$) numbers of Fock-space site energies, they generically arise from an extensive ($\mathcal{O}(N)$)
number of random terms, because $H_\text{diag}$ is a sum of $\mathcal{O}(N)$ terms each having support on only $\mathcal{O}(1)$ real-space sites. Due to these correlations, the most fundamental, albeit non-trivial, object one has to consider is the $\nh$-dimensional joint distribution of the set $\{\mathcal{E}_I\}$. Equivalently, this distribution can be considered to be that of a $N_\mathcal{H}$-dimensional vector, $\bm{\mathcal{E}}$, where each element corresponds to a particular $\mathcal{E}_I$. Crucially, as elaborated below, this vector can be represented as a sum of $\mathcal{O}(N)$ independent random vectors, where each term in the sum originates from a local term in $H_\text{diag}$. Appealing to the multivariate version of the central limit theorem, one can thus argue that the $\nh$-dimensional joint distribution must be a multivariate Gaussian.

Let us illustrate this with an example. Consider the diagonal part of the Hamiltonian to be of form
\begin{equation}
	H_\text{diag} = \sum_{\ell=1}^N\mathcal{J}_\ell\,O(\sigma^z_{\ell},\cdots,\sigma^z_{\ell+m})
	\label{eq:hdiag-zz}
\end{equation}
where each $\mathcal{J}_\ell$ is an independent $\mathcal{O}(1)$ random number, and $O(\sigma^z_{\ell},\cdots,\sigma^z_{\ell+m})$ is a local operator function made up of $\sigma^z$-operators on sites $\ell$ through $\ell+m$ with a finite 
$m$.\footnote{As an example, for the XXZ-chain with disordered fields $\{h_{i}\}$,
$\mathcal{J}_\ell\,O(\sigma^z_{\ell},\cdots,\sigma^z_{\ell+m}) = J\sigma^z_\ell\sigma^z_{\ell+1}+h_\ell\sigma^z_\ell$.} Further, let us separate $\mathcal{J}_\ell$ generally into a constant ($J$) and a disordered part, $\mathcal{J}_\ell = J + J_\ell$, where the $J_{\ell}$ ($l=1,2,..N$) are independent random variables with vanishing mean, $\braket{J_{\ell}}=0$, but finite variance $\braket{J_\ell^2} = W^2_J$.

With $H_\text{diag}$ of the general form Eq.~\eqref{eq:hdiag-zz}, the vector $\bm{\mathcal{E}}$ can be expressed as
\begin{equation}
	\bm{\mathcal{E}} = \begin{pmatrix}\mathcal{E}_1^{\pd}\\ \mathcal{E}_2^{\pd}\\ \vdots\\ \mathcal{E}_I^{\pd}\\\vdots\\ \mathcal{E}_{\nh}^{\pd}\end{pmatrix} = \sum_{\ell=1}^N \mathbf{v}_\ell=
	\sum_{\ell =1}^{N} \begin{pmatrix}\mathcal{J}_\ell O(s_{\ell}^{(1)},\cdots,s_{\ell+m}^{(1)})\\ \mathcal{J}_\ell O(s_{\ell}^{(2)},\cdots,s_{\ell+m}^{(2)})\\ \vdots\\ \mathcal{J}_\ell O(s_{\ell}^{(I)},\cdots,s_{\ell+m}^{(I)})\\\vdots\\ \mathcal{J}_\ell O(s_{\ell}^{(\nh)},\cdots,s_{\ell+m}^{(\nh)})\end{pmatrix},
	\label{eq:mvg-example}
\end{equation}
where $s_\ell^{(I)}=\pm 1$ is the $z$-component of spin at site $\ell$ in the Fock-space site $\ket{I}$.
Each of the $N$, $\nh$-dimensional vectors $\mathbf{v}_\ell$ in Eq.~\eqref{eq:mvg-example}, is independent, as required for the multivariate CLT, whence one can conclude that in the thermodynamic limit the distribution of $\bm{\mathcal{E}}$ 
over disorder realisations is a multivariate Gaussian, {i.e.}\
\begin{equation}
\mathcal{P}_{\nh}^d(\bm{\mathcal{E}}) = \frac{1}{\sqrt{(2\pi)^{\nh}\vert \mathbf{C}_d\vert}}\exp\left[-\frac{1}{2}\bm{\mathcal{E}^\prime}{}^\mathrm{T} \cdot \mathbf{C}_d^{-1}\cdot \bm{\mathcal{E}^{\prime }}\right]
\label{eq:mvg-disorder}
\end{equation}
where 
\begin{eqnarray}
\mathcal{E}_I^\prime = \mathcal{E}_I^{\pd}-\mathcal{E}_I^0&=&\sum_{\ell=1}^NJ_\ell O(s_{\ell}^{(I)},\cdots,s_{\ell+m}^{(I)})\label{eq:mvg-eiprime}\\
\mathcal{E}_I^0&=& J\sum_{\ell=1}^N O(s_{\ell}^{(I)},\cdots,s_{\ell+m}^{(I)}),
\label{eq:mvg-eizero}
\end{eqnarray}
and $\mathbf{C}_d$ is the covariance matrix characterising the distribution.
Note that as the sum in Eq.~\eqref{eq:mvg-example} has $N$ terms, the covariance matrix $\mathbf{C}_d$ scales linearly with $N$. We add further that, since any marginal distribution of a multivariate Gaussian remains of the same form,
the distribution $\mathcal{P}_{\mathcal{D}}^d$ over some $\mathcal{D}$-dimensional subspace of Fock space is also a multivariate Gaussian.

In Eq.~\eqref{eq:mvg-disorder}, the superscript $d$ in $\mathcal{P}_{\nh}^d$ (and subscript in $\mathbf{C}_d$) denotes that the distribution is over disorder realisations. The form of the distribution $\mathcal{P}_{\nh}^d$ in
Eqs.\ \eqref{eq:mvg-disorder}-\eqref{eq:mvg-eizero}, and likewise for $\mathcal{P}_{\mathcal{D}}^d$, shows that these distributions are as expected conditional distributions for the $\mathcal{E}_I$s, \emph{given} a set $\{\mathcal{E}_I^0\}$.
The $\{\mathcal{E}_I^0\}$ are by definition independent of disorder. But they are distributed over the Fock space.
One can thus consider a $\mathcal{D}$-dimensional distribution \emph{over the Fock space}, of the Fock-space site energies $\{\mathcal{E}_I^0\}$ of a $\mathcal{D}$-dimensional subspace consisting of a Fock-space site and all its neighbours lying within a certain Hamming distance (such that $\mathcal{D}\ll\nh$ for there to be a meaningful distribution). 
The simplest example is the $\mathcal{D}=1$-dimensional probability density that any Fock-space site chosen at
random from the Fock space has a particular $\mathcal{E}_{I}^{0}$~\cite{welsh2018simple}. One could also consider a patch of Fock space consisting of a Fock-space site and its $N$ nearest neighbours, and ask for the $\mathcal{D} =(N$$+$$1)$-dimensional distribution of this set of $\{\mathcal{E}_{I}^{0}\}$ over the full Fock space.

In these cases an argument can again be constructed along the lines of Eq.~\eqref{eq:mvg-example}, where instead of the independence of the $J_\ell$s, one uses the independence of the $O(s_{\ell}^{(I)},\cdots,s_{\ell+m}^{(I)})$ over the various $\ket{I}$s in disjoint $\mathcal{D}$-dimensional subspaces of Fock space (which is where the smallness of subspace dimension $\mathcal{D}$ and the finite-range nature of the models is important). Such an argument leads to the conclusion that the $\mathcal{D}$-dimensional distribution of $\{\mathcal{E}_{I}^{0}\}$ is also a multivariate Gaussian,
\begin{equation}
\mathcal{P}_{\mathcal{D}}^F(\bm{\mathcal{E}}^0) = \frac{\exp\left[-\frac{1}{2}(\bm{\mathcal{E}}^0-\overline{\mathcal{E}^0})^{\mathrm{T}} \cdot \mathbf{C}_F^{-1}\cdot (\bm{\mathcal{E}}^0-\overline{\mathcal{E}^0})\right]}{\sqrt{(2\pi)^{\mathcal{D}}\vert \mathbf{C}_F\vert}},
\label{eq:mvg-fs}
\end{equation}
where the superscript $F$ indicates that the distribution is over the Fock space, and 
$\overline{\mathcal{E}^0}=\nh^{-1}\sum_{I}\mathcal{E}_I^0$ is the mean of $\mathcal{E}_{I}^{0}$ over the Fock space.
 We consider $\overline{\mathcal{E}^0}=0$ in the following, for convenience, as it corresponds merely to an overall shift 
independent of the disorder realisation (and in fact all microscopic models considered later inherently have $\overline{\mathcal{E}^0}=0$).

As noted above, $\mathcal{P}^d_\mathcal{D}$ is a conditional distribution for the $\mathcal{E}_I$s given a set 
$\{\mathcal{E}_I^0\}$.  On the other hand, $\mathcal{P}_\mathcal{D}^F$ in Eq.~\eqref{eq:mvg-fs} is a distribution of the 
$\mathcal{E}_{I}^0$s themselves. Hence a $\mathcal{D}$-dimensional distribution over both disorder realisations and the Fock space is simply a convolution
\begin{align}
\mathcal{P}_\mathcal{D}^{\pd}(\bm{\mathcal{E}}) =& (\mathcal{P}_\mathcal{D}^d\ast\mathcal{P}^F_\mathcal{D})(\bm{\mathcal{E}})\nonumber\\
=&\frac{\exp\left[-\frac{1}{2}\bm{\mathcal{E}}^{\mathrm{T}} \cdot \mathbf{C}^{-1}\cdot \bm{\mathcal{E}}\right]}{\sqrt{(2\pi)^{\mathcal{D}}\vert \mathbf{C}\vert}},
\label{eq:mvg-disorder-fs}
\end{align}
where we use the fact that the convolution of two multivariate Gaussians is again a multivariate Gaussian whose covariance matrix is the sum of the covariance matrices of the two, {i.e.}\
\begin{equation}
\mathbf{C}=\mathbf{C}_{d,\mathcal{D}}+\mathbf{C}_F
\end{equation}
where $\mathbf{C}_{d,\mathcal{D}}$ is the $\mathcal{D}$-dimensional submatrix of $\mathbf{C}_d$ appearing in Eq.~\eqref{eq:mvg-disorder}.

An important ingredient for the above framework to hold is that, irrespective of how the $\mathcal{D}$-dimensional subspace is chosen for $\mathcal{P}_\mathcal{D}$, the elements of the covariance matrix
$\mathbf{C}=\mathbf{C}_{d,\mathcal{D}}+\mathbf{C}_F$ depend only on the Hamming distances between the Fock-space sites. We refer to this function for any given Hamming distance $r$ as $C(r)$, 
\begin{equation}
C(r) := \braket{\mathcal{E}_I^{\pd}\mathcal{E}_K^{\pd}}_{r_{IK}=r}^{\pd}\equiv\frac{2}{{\nh}\binom{N}{r}}\sum_{\substack{I,K: \\r_{IK}=r}}\braket{\mathcal{E}_I^{\pd}\mathcal{E}_K}^{\pd}
\label{eq:covariance-def}
\end{equation}
where $\nh \binom{N}{r}/2$ is the number of Fock-space site pairs with a Hamming distance $r$, and $\braket{\cdots}$ denotes an average over disorder realisations.
At this stage we leave it as an assertion that the matrix elements of $\mathbf{C}$ depend only on the Hamming distances 
(with the second equality above thus redundant), but will later show explicitly using analytic arguments and numerical results that this is indeed true in the thermodynamic limit for a 
range of microscopic systems. It is for the numerical calculations on finite-sized systems that the otherwise redundant Fock-space average in 
Eq.\ \eqref{eq:covariance-def} is practically useful.

Note that $C(r=0)$ is simply the variance of the univariate Gaussian distribution $\mathcal{P}_1(\mathcal{E})$, 
\textit{i.e.}\  $C(0)=\mu_\mathcal{E}^2$, where $\mu_\mathcal{E} \propto \sqrt{N}$ determines the width of the total density of states for the Fock-space site energies. For convenience later on, to make the $N$-dependence
of $\mu_{\mathcal{E}}$ explicit, we define $W_{\tot}$ by
\begin{equation}
\mu_{\mathcal{E}}^{\pd} ~=~ W_{\tot}^{\pd} \sqrt{N},
\label{eq:WFSdef}
\end{equation}
with $W_{\tot}$ an $\mathcal{O}(1)$ number depending on the microscopic energy scales of the model.

One further, important point should be made, regarding the thermodynamic limit of the multivariate Gaussian
distributions discussed above. We illustrate it with respect to  $\mathcal{P}_\mathcal{D}(\bm{\mathcal{E}})$, 
Eq.\ \eqref{eq:mvg-disorder-fs}. Because its covariance matrix $\mathbf{C}$ scales linearly with $N$, it is 
$\mathbf{C}/N$ whose elements remain finite in the limit $N\to \infty$; and since 
$\mu_{\mathcal{E}}^{2} \propto N$ (Eq.\ \eqref{eq:WFSdef}), we consider specifically
$\mathbf{C}_\mathbf{x} =\mathbf{C}/\mu_{\mathcal{E}}^{2}$. 
Rescaling the exponential in Eq.\ \eqref{eq:mvg-disorder-fs} then shows that it is the
distribution of the $x_{I}=\mathcal{E}_{I}/\mu_{\mathcal{E}}$, rather than $\mathcal{E}_{I}$ \emph{per se},
which has a well-defined thermodynamic limit; given by
\begin{equation}
\mathcal{P}_\mathcal{D}^{\pd}(\mathbf{x}) =\frac{1}{\sqrt{(2\pi)^{\mathcal{D}}\vert \mathbf{C}_\mathbf{x}\vert}}
\exp\left[-\frac{1}{2}\mathbf{x}^{\mathrm{T}} \cdot \mathbf{C}_\mathbf{x}^{-1}\cdot \mathbf{x}\right]
\label{eq:mvg-disorder-fs_rescaled}
\end{equation}
(with $\mathbf{x}^{\mathrm{T}}$ a row vector whose elements are  $x_{I} =\mathcal{E}_{I}/\mu_{\mathcal{E}}$).
The many-body  site energies must in other words be rescaled $\propto \mathcal{E}_{I}/\sqrt{N}$
to ensure a well-defined thermodynamic limit; a fact which also plays an important role in the self-consistent mean-field
theory of Sec.\ \ref{sec:fockspacelocalisation}.

In the following sections it is the precise behaviour of the correlations in the Fock-space site energies, encoded in 
$C(r)/C(0)=C(r)/\mu_{\mathcal{E}}^{2}$ Eq.~\eqref{eq:covariance-def}, that will take centre stage in understanding 
when -- and when not -- a many-body localised phase can be stabilised.
Let us now summarise the minimal characteristics of Fock-space graphs with which we will be concerned:
\begin{itemize}
	\item The average connectivity on the Fock-space graph scales linearly with $N$.
		\item The distribution, over both disorder realisations and the Fock-space, 
		 of a small subspace consisting of Fock-space sites restricted by the Hamming distance, is described by a multivariate 
		  Gaussian.
	\item The elements of the covariance matrix of this multivariate Gaussian scale linearly with $N$.
	\item The  covariance matrix elements depend only on the Hamming distances between the Fock-space sites.
\end{itemize}

\begin{figure}
\includegraphics[width=\columnwidth]{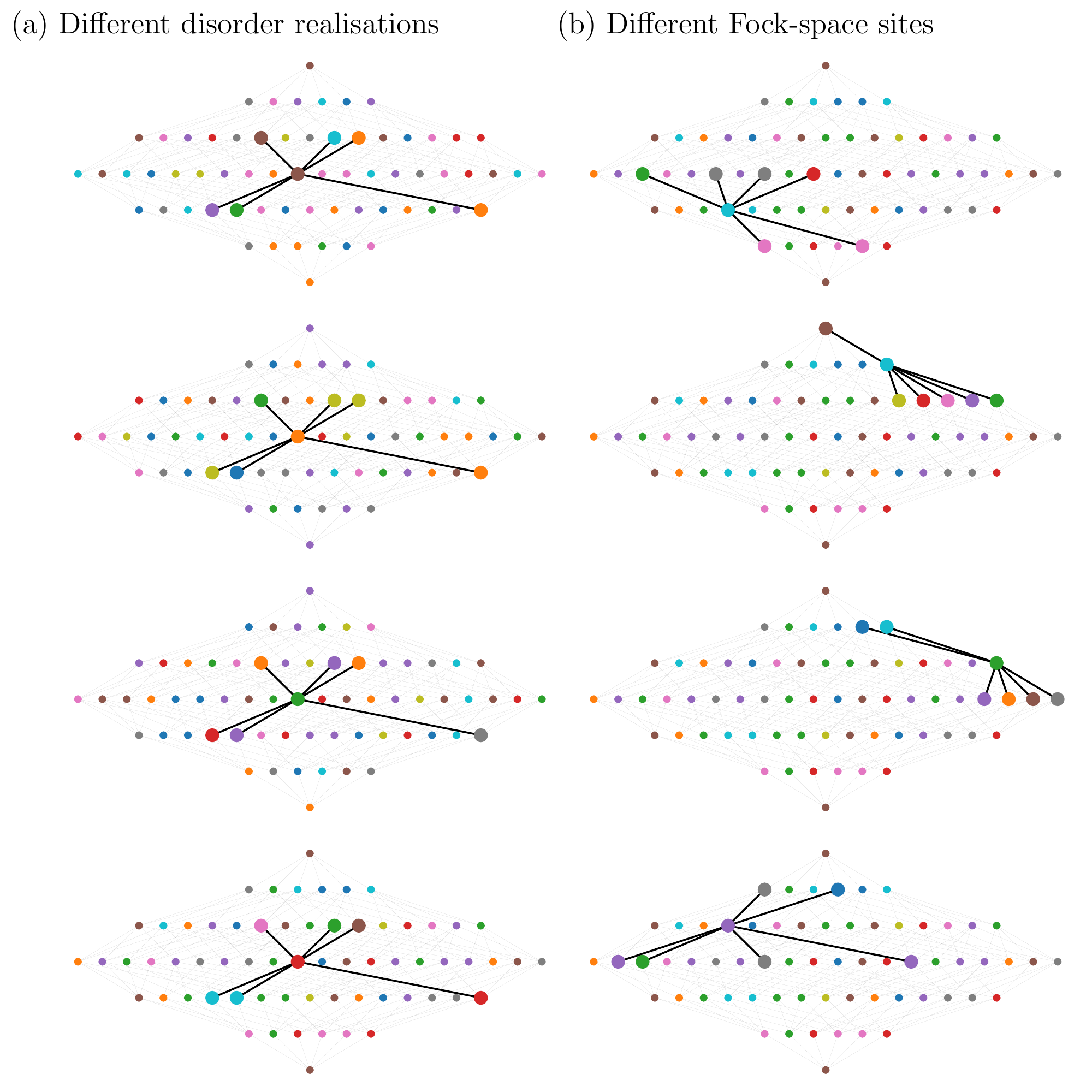}
\caption{
Schematic of the Fock-space graph and the two types of distribution: over disorder realisations (a), and over the Fock space (b). Each Fock-space site on the graph corresponds to a spin-configuration, with random colours denoting the random (but correlated) on-site energies. The links are generated by $H_\text{1}$ in Eq.~\eqref{eq:ham-offdiag}. In the set of panels (a), each row corresponds to a different disorder realisation of the random energies. Hence keeping a fixed Fock-space site and its neighbours (highlighted with larger circles) allows one to construct the distribution over disorder realisations, $\mathcal{P}_{N+1}^d$. Instead, as illustrated in the set of panels (b), one could move the Fock-space site and its neighbours over the Fock space, while keeping a fixed disorder realisation (as shown by the same set of colours), to generate the distribution $\mathcal{P}_{N+1}^F$.}
\label{fig:kinds-of-distributions}
\end{figure}

For concreteness, let us illustrate some of the above ideas using the Fock-space graph wherein the links are generated by 
$H_\text{1}$ in Eq.~\eqref{eq:ham-offdiag}. Consider the $\mathcal{D}$-dimensional subspace to be made up of a Fock-space site, denoted by $\ket{I_0}$, and its nearest neighbours (Hamming distance of 1) $\ket{I_\alpha}$, such that $\alpha=1,2,\cdots N$. The distribution $\mathcal{P}^d_{N+1}$ can then be understood as fixing the Fock-space sites and 
 sampling over many disorder realisations; this is illustrated schematically in Fig.~\ref{fig:kinds-of-distributions}(a) 
(for a tiny Fock space of dimension $\nh =2^{6}$). Instead, as illustrated in Fig.~\ref{fig:kinds-of-distributions}(b), to generate the distribution $\mathcal{P}_{N+1}^F$ one can fix the disorder realisation  -- or equivalently look at the disorder independent $\mathcal{E}_{I}^{0}$s contributing  to the $\mathcal{E}_{I}$s -- and move the Fock-space site together with its neighbours over the Fock space (thereby sampling many `patches' of size $\mathcal{D}=N$$+$$1$).
Since $\mathcal{D}/\nh =2^{-N}(N+1)$ vanishes in the thermodynamic limit, note that only disjoint $\mathcal{D}$-dimensional
patches then contribute to this distribution (as overlapping patches form a set of measure zero).

Note that the structure of the Fock-space stipulates the Hamming distances $r_{I_0 I_\alpha}=1$ and $r_{I_\alpha I_{\alpha^\prime}}=2(1-\delta_{\alpha\alpha^\prime})$. Hence, arranging the vector $\bm{\mathcal{E}}=(\mathcal{E}_{I_0},\mathcal{E}_{I_1},\cdots,\mathcal{E}_{I_N})^\mathrm{T}$, the joint distribution, $\mathcal{P}_{N+1}(\bm{\mathcal{E}})$, is of the form Eq.~\eqref{eq:mvg-disorder-fs}; with the $(N$$+$$1)$-dimensional covariance matrix 
\begin{equation}
\mathbf{C} = \begin{bmatrix}
C(0)&C(1)&\cdots&\cdots&C(1)\\
C(1)&C(0)&C(2)&\cdots&C(2)\\
\vdots&C(2)&C(0)&\ddots&\vdots\\
\vdots&\vdots&\ddots&\ddots&C(2)\\
C(1)&C(2)&\cdots&C(2)&C(0)
\end{bmatrix}
\label{eq:c-matrix-tfi}
\end{equation}
consisting of $C(0)$ along the diagonals and, for the off-diagonal elements, $C(1)$ alone in the first row and column, and
$C(2)$ for all other entries.


\section{Local propagators in Fock space and localisation criterion 
\label{sec:fockspacelocalisation}}

In analysing localisation on the Fock space, we consider the Fourier transform, $G_{I}(\w)$, of the local Fock-space propagator $G_I(t)=-i\Theta(t)\braket{I|e^{-iHt}|I}$. This can be expressed as  
\begin{eqnarray}
G_I^{\pd}(\omega) &=& \braket{I|(\omega + i\eta - H)^{-1}|I} \nonumber \\
&=&[\omega^+ -\mathcal{E}_I^{\pd}-\Sigma_I^{\pd}(\omega)]^{-1},
\label{eq:fs-propagator}
\end{eqnarray}
where $\omega^+=\omega+i\eta$ with $\eta=0^+$ the regulator.  The self-energy  for Fock-space site $I$ is 
separated as $\Sigma_I = X_I -i\Delta_I$ into real and imaginary parts. 
Our natural attention will centre on $\Delta_I(\omega)$, which in physical terms is the inverse lifetime (or loss rate) 
associated with the decay of probability from $|I\rangle$ into states of energy $\w$.
The basic approach here follows that of Ref.\ \cite{logan2019many,roy2019self}, where further discussion is given 
of the points now briefly reprised.

In the context of single-particle localisation, for a delocalised state in the thermodynamic limit, $\Delta_I$ 
takes a finite value with unit probability (over an ensemble of disorder realisations).  For a localised state, 
by contrast, it vanishes with unit probability, signifying an infinite lifetime for the state.
In the many-body case however, $\Delta_I$ itself is not an appropriate diagnostic of many-body localisation, 
due to its non-trivial scaling with $N$. In the many-body delocalised phase, $\Delta_I$ diverges 
as $\sqrt{N}$ with probability unity \cite{logan2019many}. It is thus the rescaled quantity 
$\Delta_I/\sqrt{N}$ that admits a finite value in the thermodynamic limit.
In the many-body localised phase on the other hand, $\Delta_I$ vanishes with unit probability; specifically 
$\Delta_I \propto \eta \rightarrow 0^{+}$ such that $\Delta_{I}/\eta$ is finite
(or equivalently $\tilde{\Delta}_{I}/\tilde{\eta}$, where $\tilde{\Delta}_{I}\propto \Delta_{I}/\sqrt{N}$ and
likewise for $\tilde{\eta}$).

The rescaling of many-body energy scales with $\sqrt{N}$ is an important constituent of the theory,
as already seen in the discussion of Sec.\ \ref{sec:FSgraph} (Eq.\ \eqref{eq:mvg-disorder-fs_rescaled})
regarding the thermodynamic limit of the multivariate Gaussian $\P_{\mathcal{D}}$;
whence some further brief elaboration is in order  (with details given in \cite{logan2019many,roy2019self}).
This rescaling is also required
by the  $N$-dependent scaling of the total density of states, and of the eigenstate amplitudes on the Fock-space. Note that the local density of states, $D_I(\omega)$, for Fock-space site $I$ is given by the imaginary part of the local
 propagator, $D_I(\omega)=-\pi^{-1}\Im[{G_I(\omega)}] = \sum_n \vert A_{nI}\vert^2\delta(\omega-E_n)$, where 
$ \vert A_{nI}\vert^2=\vert\braket{E_n|I}\vert^2$ is the amplitude on Fock-space site $I$ of the eigenstate $\ket{E_n}$ with eigenvalue $E_n$. Deep in the delocalised phase, $\vert A_{nI}\vert^2$ scales typically as $N_\mathcal{H}^{-1}$ irrespective of the specific $n$ and $I$, and thus $D_I(\omega)$ is equal to the normalised \emph{total} density of states. The latter is a Gaussian with a standard deviation $\mu_E\propto \sqrt{N}$~\cite{welsh2018simple,logan2019many}. Thus, 
$D_I(\omega)\propto\mu_E^{-1}\propto 1/\sqrt{N}$ for all $\omega$.

The utility of studying $D_I(\omega)$ in the context of localisation lies in part in the fact that in the delocalised phase, $D_I(\omega)$ forms a continuum in $\omega$. This reflects that any site $I$ has an overlap with 
essentially all delocalised eigenstates at all energies $\omega$ (the number of which is $\nh$, diverging exponentially 
in $N$). On the other hand, in a one-body context, $D_I(\omega)$ in a localised phase
is pure point-like, which signifies that a site $I$ -- in this case a single real-space site -- overlaps only an $\mathcal{O}(1)$ number of eigenstates at specific energies. The situation for a many-body localised phase is somewhat more subtle. Akin to the delocalised phase, a Fock-space site $I$ again has support on an 
exponentially diverging (in $N$) number of eigenstates, and thus $D_I(\omega)$  strictly speaking forms a 
continuum in $\omega$. But in contrast to the delocalised phase, the \emph{fraction} of eigenstates on which a Fock-space site has support vanishes in the thermodynamic limit; as manifest in the typical number of such states scaling as $N_\mathcal{H}^\alpha$ with $\alpha<1$.

To resolve this characteristic of the many-body localised phase, one thus needs to study the local (on Fock space) density of states on scales relative to that of the delocalised phase. As noted above, $D_I(\omega)$ in the delocalised phase scales as $1/\sqrt{N}$. Hence it is $\tilde{D}_I(\w) =\sqrt{N}D_I(\w)$ which both admits a 
thermodynamic limit and exposes the distinction between the localised and delocalised phases in the many-body case.

Related to the above, a simple but suggestive physical argument can also be  given
as to why $\Delta_I$ scales as $\sqrt{N}$ in a delocalised phase. Physically, $\Delta_I(\omega)$ is 
proportional to 
the rate of loss of probability from a Fock-space site $I$ to other sites in states of energy $\omega$. Note that any Fock-space site has $\mathcal{O}(N)$ neighbours, so there are $\mathcal{O}(N)$  channels for the probability loss. At the same time, the density of states at energy $\omega$ scales as $1/\sqrt{N}$. A simple Fermi golden rule-like argument then implies that
the effective rate of loss is proportional to the product of the two, and hence scales as $\sqrt{N}$. This again shows that it is the rescaled  $\Delta_I/\sqrt{N}$ which is the appropriate quantity to study in the thermodynamic limit.


\subsection{Self-consistent theory}
\label{subsection:selfcontheory}

The self-energy $\Sigma_I(\omega)$ can be recast using the Feenberg renormalised perturbation 
series~\cite{feenberg1948note,Economoubook}, 
\begin{eqnarray}
\Sigma_I^{\pd}(\omega) &=& \sum_{K} T_{IK}^2 G_K^{\pd}(\omega) + \cdots \nonumber\\
&=&\sum_{K}\frac{T_{IK}^2}{\omega^+-\mathcal{E}_K^{\pd}-\Sigma_K^{\pd}(\omega)}+\cdots,
\label{eq:sigma-series}
\end{eqnarray} 
where in this work, as in Ref.\ \cite{logan2019many,roy2019self}, we truncate at the renormalised second order indicated.
In particular, since we work with the off-diagonal part of the Hamiltonian given by Eq.~\eqref{eq:ham-offdiag}, for a Fock-space site $I_0$ the imaginary part of the self-energy at the renormalised second-order level can be written as
$\Delta_{I_{0}}(\w)=\pi\Gamma^{2}\sum_{\alpha} D_{I_{\alpha}}(\w)$, i.e.\
\begin{equation}
\Delta_{I_0}^{\pd}(\omega) = \Gamma^2\sum_{\alpha=1}^N\frac{\eta+\Delta^{\pd}_{I_\alpha}(\omega)}{[\w-\mathcal{E}^{\pd}_{I_\alpha}-X_{I_\alpha}^{\pd}(\omega)]^2 + [\eta+\Delta^{\pd}_{I_\alpha}(\omega)]^2}
\label{eq:delta-I0}
\end{equation}
where,  as before, $\{I_\alpha|\,\alpha=1,\cdots,N\}$ is the set of neighbours of $I_0$ on the Fock space.

A self-consistent mean-field treatment of $\Delta_{I_0}$ starting from Eq.~\eqref{eq:delta-I0} consists of 
three essential steps \cite{logan2019many,roy2019self}.
First,  $\Delta_{I_\alpha}$ and $X_{I_\alpha}$ on the right-hand side of the equation are replaced by their typical values
$\Delta_{\typ}$ and $X_{\typ}$; this constitutes the mean-field approximation of the theory.
Second, the probability distribution for $\Delta_I$, which itself depends on $\Delta_\typ$, is then obtained by 
averaging over the $(N+1)$-dimensional joint distribution of the $\{\mathcal{E}_{I_\alpha}|\,\alpha=0,1,\cdots,N\}$. 
We denote this distribution by $F_\Delta(\Delta,\Delta_\typ)$. Finally, self-consistency is imposed by calculating 
$\Delta_\typ$ from this distribution and equating it to the $\Delta_\typ$ that parametrically enters the distribution, via
\begin{equation}
\Delta_\typ^{\pd} = \exp\left[\int_0^\infty d\Delta\,\ln(\Delta)\, F_\Delta^{\pd}(\Delta,\Delta_\typ^{\pd})\right].
\label{eq:self-consistency-delta}
\end{equation}
In the following, we focus solely on the centre of the spectrum $\omega=0$, for which $X_{\typ}(\omega=0) =0$  by symmetry.

In the many-body localised phase, where $\Delta_I\propto \eta$ and $\eta=0^+$, it is of course $y_I=\Delta_I/\eta$ which is the relevant quantity to study. From Eq.~\eqref{eq:delta-I0}, it can be expressed as
\begin{equation}
y_{I_0^{\pd}}^{\pd} = \kappa\sum_{\alpha=1}^{N}\frac{1}{\mathcal{E}_{I_\alpha}^2}; ~~~~~~:~ \kappa = \Gamma^2(1+y_\typ^{\pd}),
\label{eq:yI0-strong}
\end{equation}
where we have set $\omega=0$. The distribution of $y_{I}$ can thus be obtained as 
\begin{equation}
F_y(y,y_\typ^{\pd}) = \int [d^{N+1}\bm{\mathcal{E}}]\,\mathcal{P}_{N+1}^{\pd}(\bm{\mathcal{E}})\delta\left(y-\kappa\sum_{\alpha=1}^{N}\frac{1}{\mathcal{E}_{I_\alpha}^2}\right).
\label{eq:Fy}
\end{equation}
It is whether or not the distribution obtained from Eq.~\eqref{eq:Fy}  satisfies the self-consistency condition 
Eq.~\eqref{eq:self-consistency-delta} that constitutes the criterion for localisation.
If self-consistency is indeed satisfied for a finite range of $\Gamma$ and $W_\tot=\mu_\mathcal{E}/\sqrt{N}$
(each of which are $\mathcal{O}(1)$ quantities), then a many-body localised phase can be stabilised.
If by contrast the self-consistency breaks down for any non-zero value of $\Gamma$, then a localised phase
cannot be stabilised and the system  is always delocalised.

The distribution $\mathcal{P}_{N+1}(\bm{\mathcal{E}})$ is an $(N+1)$-dimensional Gaussian with covariance matrix 
$\mathbf{C}$ given 
by Eq.~\eqref{eq:c-matrix-tfi}. However, as discussed in Sec.\ \ref{sec:FSgraph},
it is the rescaled variables $\mathcal{E}_I/\sqrt{N}$ that 
are required. Since $\mu_\mathcal{E}=W_\tot\sqrt{N}$ also scales as $\sqrt{N}$, we work in practice with the 
rescaled variables $x_I = \mathcal{E}_I/\mu_\mathcal{E}$, such that the $(N+1)$-dimensional Gaussian distribution of 
$\{\mathbf{x}\}$ (Eq.\ \eqref{eq:mvg-disorder-fs_rescaled})
has a covariance matrix $\mathbf{C}_\mathbf{x}$ given by
\begin{equation}
\mathbf{C}_\mathbf{x}=\frac{1}{\mu_\mathcal{E}^2}\mathbf{C} = \begin{bmatrix}
1&\rho_1&\cdots&\cdots&\rho_1\\
\rho_1&1&\rho_2&\cdots&\rho_2\\
\vdots&\rho_2&1&\ddots&\vdots\\
\vdots&\vdots&\ddots&\ddots&\rho_2\\
\rho_1&\rho_2&\cdots&\rho_2&1
\end{bmatrix}
\label{eq:cx-matrix-tfi}
\end{equation}
where $\rho_r = C(r)/C(0) = C(r)/\mu_\mathcal{E}^2$. Note that, being a covariance matrix, the matrix 
$\mathbf{C}_\mathbf{x}$ is necessarily positive semidefinite, which places constraints on $\rho_1$ and $\rho_2$,
\begin{equation}
0\le\vert\rho_1\vert,\rho_2\le1~; ~~~\rho_1^2 \le \rho_2 + \frac{1}{N}(1-\rho_2).
\label{eq:rho-constraints}
\end{equation}

In terms of the rescaled variables, $x_I$, the distribution $F_y$ from Eq.~\eqref{eq:Fy} can be rewritten as
\begin{equation}
F_y(y,y_\typ^{\pd}) = \int [d^{N+1}\mathbf{x}]\,\mathcal{P}_{N+1}^{\pd}(\mathbf{x})\delta\left(y-\frac{\kappa}{\mu_\mathcal{E}^2}\sum_{\alpha=1}^{N}\frac{1}{x_{I_\alpha}^2}\right).
\label{eq:Fy-rescaled}
\end{equation} 
Importantly, the argument of the $\delta$-function here does not depend on $x_{I_0}$. 
It can thus be integrated out to give
\begin{equation}
\begin{split}
&F_y^{\pd}(y,y_\typ^{\pd}) =
\\
& \int \big[\prod_{\alpha=1}^N d x_{I_\alpha}^{\pd}\big]\,\mathcal{P}_{N}^{\pd}(\{x^{\pd}_{I_\alpha}\})\delta\left(y-\frac{\kappa}{\mu_\mathcal{E}^2}\sum_{\alpha=1}^{N}\frac{1}{x_{I_\alpha}^2}\right)
\label{eq:Fy-x0-integ}
\end{split}
\end{equation}
where we have made explicit the arguments in the integral measure. 
Recalling that any marginal distribution of a multivariate Gaussian is also a multivariate Gaussian, 
$\mathcal{P}_{N}(\{x_{I_\alpha}\})$ in Eq.\ \eqref{eq:Fy-x0-integ} is an $N$-dimensional Gaussian distribution.
Its covariance matrix depends solely on $\rho_2$, because all Fock-space sites in the set $\{I_\alpha\}$ have a
mutual Hamming distance of 2. In fact the covariance matrix is the same as in Eq.~\eqref{eq:cx-matrix-tfi}, but 
with the first row and column deleted.
Hence the distribution $F_y$ depends only on $\rho_2$ and not on $\rho_1$.
This provides us a free choice of $\rho_1$ to work with for analytic convenience 
(subject to satisfaction of the constraints Eq.~\eqref{eq:rho-constraints}), knowing that the results are 
independent of it. Interestingly, such a choice is provided by $\rho_1^2=\rho_2$, in which case the form of 
the covariance matrix makes computation of the distribution $F_y$ amenable to an analytic treatment.
In addition, as will be shown in Sec.~\ref{sec:microscopic},  a large class of microscopic models with local interactions satisfy $\rho_{1}^{2}=\rho_{2}$ modulo $\mathcal{O}(N^{-2})$ corrections.

In the following we will present analytic results for the case $\rho_1^2=\rho_2$;
emphasising again that the choice is merely an analytic device and that the results hold generally for 
any $\rho_2$. Any result that we derive in terms of $\rho_1$ and $\rho_{2}$ could thus be generalised 
in terms of $\rho_2$ alone, simply by setting $\rho_1^2\rightarrow\rho_2$.


\subsection{Analytic results}
\label{subsec:analyticresults}

To expose the analytic convenience of the choice $\rho_1^2=\rho_2$, first note that the distribution $\mathcal{P}_{N+1}(\mathbf{x})$ can be expressed as
\begin{equation}
\mathcal{P}_{N+1}^{\pd}(\mathbf{x}) = \mathcal{P}_1^{\pd}(x_{I_0}^{\pd})\tilde{\mathcal{P}}_{N}^{\pd}(\{x^{\pd}_{I_\alpha}\}\vert\, x_{I_0}^{\pd}).
\end{equation}
The first factor here is the univariate normal distribution for a single rescaled Fock-space site energy, and the second is the $N$-dimensional conditional distribution of the rescaled Fock-space site energies of the neighbours of $I_0$, given $x_{I_0}$.
For the case $\rho_2=\rho_1^2$, the conditional distribution splits up into a product of $N$ univariate conditional distributions as~\footnote{Note that this decomposition into a product of univariate conditional distributions does \emph{not} imply independence of the site energies of any pair of $I_{0}$'s neighbours, $I_\alpha$ and $I_{\alpha^\prime}$ 
(which have $r_{I_\alpha I_{\alpha^\prime}}=2$). By virtue of the fact that both of them are correlated to $I_0$, they are also correlated to each other, and the correlations between the pairs are slaved to each other.}
\begin{equation}
\mathcal{P}_{N+1}^{\pd}(\mathbf{x}) = \mathcal{P}_1^{\pd}(x_{I_0}^{\pd})\prod_{\alpha=1}^N\tilde{\mathcal{P}}_{1}^{\pd}(x_{I_\alpha}^{\pd}\vert\, x_{I_0}^{\pd}),
\label{eq:pairwise-product}
\end{equation}
with
\begin{equation}
\mathcal{P}_1^{\pd} \equiv \mathcal{N}(0,1), ~~~
\tilde{\mathcal{P}}_{1}^{\pd}(x^{\pd}_{I_\alpha}\vert\, x_{I_0}^{\pd})\equiv\mathcal{N}(\rho_1^{\pd}x_{I_0}^{\pd},1-\rho_1^2),
\label{eq:univariate-gaussian}
\end{equation}
where $\mathcal{N}(\nu,\sigma^2)$ denotes a normal distribution with mean $\nu$ and variance $\sigma^2$.

As $y_I$ is the sum of many random variables, it is more convenient to work with its generating function, $\braket{e^{iky}}=\int dy\, e^{iky}F_y(y):= F_k(k)$. Using the decomposition of $\mathcal{P}_{N+1}$ (Eq.~\eqref{eq:pairwise-product})
in Eq.\ \eqref{eq:Fy-rescaled}, this can be expressed as
\begin{align}
F_k^{\pd}(k) = \int dx_{I_0}^{\pd}\,&\mathcal{P}_1^{\pd}(x_{I_0}^{\pd})\times\nonumber\\&\left[\int dx_{I_\alpha}^{\pd}\,\tilde{\mathcal{P}}_1^{\pd}(x_{I_\alpha}^{\pd}\vert x_{I_0}^{\pd})\exp\left({\frac{ik\kappa}{\mu_\mathcal{E}^2x_{I_\alpha}^2}}\right)\right]^N.
\label{eq:Fk}
\end{align}
In the following, we will explicitly compute $F_k$ and obtain the distribution $F_y(y,y_\typ)$ via a Fourier transform;
leading to a general criterion that must be satisfied by the covariance matrix $\mathbf{C}_\mathbf{x}$ 
in order for a many-body localised phase to be stable.


\subsubsection{Maximally correlated limit 
\label{subsec:maximally}}
Consider first the limiting case where $\rho_r \to 1$ for any finite $r$ as $N \to \infty$. We call this the \emph{maximally correlated limit}, since $\vert\rho_{r}\vert\le 1$ for $\mathbf{C}_\mathbf{x}$ to be a valid covariance matrix.
Thus $\rho_1=1=\rho_2$ is the case where the Fock-space site energies are as strongly correlated 
as possible; in fact they are completely slaved to each other, as reflected in the conditional distribution approaching a $\delta$-function,
\begin{equation}
\lim_{\rho_1,\rho_2\to1}\tilde{\mathcal{P}}_{1}^{\pd}(x_{I_\alpha}^{\pd}\vert\, x_{I_0}^{\pd}) = \delta(x_{I_\alpha}^{\pd}- x_{I_0}^{\pd}).
\label{eq:conditional-strong}
\end{equation}
Consequently, $F_k(k)$ from Eq.~\eqref{eq:Fk} takes the form
\begin{equation}
F_k^{\pd}(k) = \int dx_{I_0}^{\pd}\, \mathcal{P}_1^{\pd}(x_{I_0}^{\pd})\exp\left(\frac{ik\kappa N}{\mu_\mathcal{E}^2x_{I_0}^2}\right).
\label{eq:Fk-strong}
\end{equation}
With $\mathcal{P}_1^{\pd} \equiv \mathcal{N}(0,1)$ the standard normal distribution, and recalling 
$\mu_{\mathcal{E}}^{2} =W_{\tot}^{2}N$ (Eq.\ \eqref{eq:WFSdef}), this gives
\begin{equation}
F_k^{\pd}(k) =
\exp\left[-[1-i\:\mathrm{sgn}(k)]\left(\frac{\vert k\vert \kappa}{W_{\tot}^{2}}\right)^\frac{1}{2}\right],
\label{eq:Fk-stronga}
\end{equation}
a Fourier transform of which yields the desired distribution
\begin{equation}
F_y^{\pd}(y,y_\typ^{\pd})=\sqrt{\frac{\kappa}{2W_{\tot}^{2}}}~~y^{-3/2}\exp\left(-\frac{\kappa }{2W_{\tot}^{2}~y}\right),
\label{eq:Fy-strong}
\end{equation}
where $\kappa=\Gamma^2(1+y_\typ)$.
The self-consistency condition of Eq.~\eqref{eq:self-consistency-delta} in terms of $y=\Delta/\eta$ is 
$\ln (y_{\typ}) =\int_{0}^{\infty}dy~\ln(y)F_{y}(y,y_{\typ})$. 
Imposing this on Eq.\ \eqref{eq:Fy-strong} yields
$\ln(y_{\typ})=\ln\big[(2\Gamma^{2}/W_{\tot}^{2})(1+y_{\typ})\big] +\gamma$ (with $\gamma$
the Euler-Mascheroni constant), and hence
\begin{equation}
y_\typ^{\pd}=\frac{2\Gamma^2e^\gamma}{W^2_\tot-2\Gamma^2e^\gamma}.
\label{eq:ytyp-strong}
\end{equation}
Importantly, note that $y_\typ\ge0$ necessarily, since $\Delta_\typ$ is the typical rate of loss of probability from a Fock-space site and is thus non-negative. From Eq.\ \eqref{eq:ytyp-strong}, this is self-consistently possible
only if $W_{\tot}^2>2\Gamma^2e^\gamma$, showing that there indeed exists a finite range of $\Gamma/W_\tot$ where the many-body localised phase is stable. The breakdown of the self-consistency condition at $W_\tot=\sqrt{2}\Gamma e^{\gamma/2}$ corresponds to the critical point for the many-body localisation-delocalisation transition.

An analogous calculation for the stability of the delocalised phase can also be performed, centring on 
the distribution $F_{\tilde{\Delta}}(\tilde{\Delta},\tilde{\Delta}_{\typ})$ (where 
$\tilde{\Delta} =\Delta_{I}/\mu_{\mathcal{E}} \propto \Delta_{I}/\sqrt{N}$), with self-consistency again enforced 
via Eq.\ \eqref{eq:self-consistency-delta}.
Details of this calculation can be found in Ref.\ ~\cite{logan2019many,roy2019self}. We do not repeat them here, 
but the important point is that the corresponding self-consistency condition breaks down at precisely the same value of 
$\Gamma/W_\tot$ found above on approaching the transition from the localised side; thus confirming the
mean-field critical point as $W_\tot=\sqrt{2}\Gamma e^{\gamma/2}$.

The long-tailed ($\propto y^{-3/2}$) L\'evy distribution obtained for $F_y$ in Eq.~\eqref{eq:Fy-strong} is known to be 
characteristic of the many-body localised phase~\cite{logan2019many,roy2019self}. It is both interesting and reasuring to note that  this distribution, as well as the localisation criterion,  are formally the same as those obtained for disordered quantum systems with local interactions,\footnote{A Hamiltonian is defined to be local if it comprises a sum of terms each of which acts on no more than $p$ sites, with $p/N \to 0$ as $N \to \infty$.} both short and long-ranged~\cite{logan2019many,roy2019self}.
In these previous works, similar to the current one, the analysis was performed after rescaling all energy scales 
with $\sqrt{N}$. It was also identified that for two Fock-space sites $I$ and $K$ which are separated by a finite Hamming distance $r_{IK}$, the difference in Fock-space site energies typically scales as $\sqrt{r_{IK}}$ (and is proportional to the $\mathcal{O}(1)$ microscopic coupling constants in $H_\mathrm{diag}$). The $\sqrt{r}$ factor can be understood simply from the fact that two Fock-space sites separated by a Hamming distance $r$ -- and thus with $r$ spins different --
differ in their site energies by a sum of $r$ independent $\mathcal{O}(1)$ random numbers. Thus in terms of the rescaled Fock space site energies $\tilde{\mathcal{E}}_I = \mathcal{E}_I/\sqrt{N}$, the difference scales as $\sqrt{r_{IK}/N}$ and hence vanishes as $N\to\infty$, \textit{i.e.} $\tilde{\mathcal{E}_I}-\tilde{\mathcal{E}_K}\to 0$. 
From Eq.~\eqref{eq:yI0-strong} one thus has $y_{I_{0}} =(\kappa/N)\sum_{\alpha}\tilde{\mathcal{E}}^{-2}_{I_{\alpha}}
= \kappa/\tilde{\mathcal{E}}_{I_0}^2$.
Using $\mathcal{P}_1(\tilde{\mathcal{E}}_{I_0}) \equiv \mathcal{N}(0,W^2_\tot)$ then yields
exactly the same distribution  $F_{y}(y,y_{\typ})$  as Eq.~\eqref{eq:Fy-strong}.

The maximally correlated limit of our general analysis -- $|\rho_r|\to 1$ for finite $r$ in the thermodynamic limit -- captures this precise condition that the Fock-space site energies for two sites separated by finite Hamming distance have a covariance $C(r)$ such that the rescaled site energies coincide on the scale of the width of the density of states 
($\propto\sqrt{N}$). The treatment presented in this section  reveals how the covariance matrix of the joint distribution encoding this effect leads to a stable localised phase; and underscores the importance of rescaling the many-body energy scales with $\sqrt{N}$, which is crucial for $\rho_r$ to take its limiting value of 1.
 
Finally here, note that to obtain Eq.~\eqref{eq:Fk-stronga} from Eq.~\eqref{eq:Fk-strong} we used that $\P_1$ is a Gaussian. The essential result is not however limited to that case. One can relax the requirement on $\P_1$, and yet show that a localised phase remains stable in the maximally correlated limit; we show this explicitly in  Appendix~\ref{app:nongaussian}.


\subsubsection{Random energy limit}
\label{subsubsec:QREMlimit}

Let us turn now to the diametrically opposite limit,  namely that of uncorrelated and independent Fock-space site energies. Note that such a scenario cannot arise from a local model such as that in Eq.~\eqref{eq:hdiag-zz}, since one necessarily requires  $2^N$ independent random energies to be assigned to the Fock-space sites; in our case drawn from $\mathcal{N}(0,W_{\tot}^2N)$. The scaling of the variance with $N$ ensures that the model mimics a many-body system with regard to the density of states. The model with such purely random Fock-space site energies and Fock-space hoppings generated by Eq.~\eqref{eq:ham-offdiag}, has the eponymous title of \emph{quantum random energy model} (QREM). It has been shown to have delocalised eigenstates in the middle of the spectrum even for arbitrarily small values of $\Gamma$~\cite{laumann2014many,baldwin2016manybody}.~\footnote{Refs.~\cite{laumann2014many,baldwin2016manybody} also argue that there exists a mobility edge in the QREM at finite energy densities, such that states with energies further than $\sim \Gamma N$ away from the band centre are many-body localised. However, since the width of the density of states scales as 
$\sqrt{N}$, the fraction of such states vanishes exponentially in $N$; so for all intents and purposes the QREM is delocalised at any non-zero value of $\Gamma$ in the thermodynamic limit.}

We now show how our self-consistent theory, developed generally in terms of the covariance matrix, encompasses this result. The case of all Fock-space site energies being independent means $C(r)=0=\rho_r$ for any non-zero $r$, and hence
\begin{equation}
\mathcal{P}_{N+1}^{\pd}(\mathbf{x}) = \prod_{\alpha=0}^N \mathcal{P}_1^{\pd}(x_{I_\alpha}^{\pd}),
\label{eq:product-P-qrem}
\end{equation}
where $\mathcal{P}_1(x_{I_\alpha})$ is the  usual normal distribution. Using Eq.~\eqref{eq:product-P-qrem} in
Eq.\ \ref{eq:Fy} (or \eqref{eq:Fk}),  one obtains for the QREM
\begin{equation}
F_k^{\pd}(k)=\exp\left[-[1-i\:\mathrm{sgn}(k)]\left(\frac{\vert k\vert N\kappa}{W_{\tot}^{2}}\right)^\frac{1}{2}\right].
\label{eq:Fk-qrem}
\end{equation}
This is of the same form as Eq.~\eqref{eq:Fk-stronga} obtained in the maximally correlated limit but -- crucially -- with $\kappa$ replaced with $\kappa N$. 
The origin of this extra factor of $N$ is readily understood, on noting from Eq.~\eqref{eq:yI0-strong}
that $y_{I_{0}} =(\kappa/N)\sum_{\alpha =1}^{N}\tilde{\mathcal{E}}^{-2}_{I_{\alpha}}$ generally (recall
that $\tilde{\mathcal{E}}_{I}:=\mathcal{E}_{I}/\sqrt{N}$).
In the maximally correlated limit, as above, this sum reduces to a single term,
$y_{I_{0}}=\kappa/\tilde{\mathcal{E}}_{I_0}^2$, due to the maximal correlation between the neighbouring 
Fock-space site energies. For the QREM by contrast the sum  contains \emph{N independent} terms, because there 
are no correlations between the Fock-space site energies.

The resultant distribution $F_y(y,y_\typ)$ for the QREM is also a L\'evy as in Eq.~\eqref{eq:Fy-strong}, but again with $\kappa$ replaced with $\kappa N$. The self-consistency condition then yields 
(\textit{cf} Eq.\ \eqref{eq:ytyp-strong})
\begin{equation}
y_\typ^{\pd}=\frac{2\Gamma^2Ne^\gamma}{W^2_\tot-2\Gamma^2Ne^\gamma}
\label{eq:ytyp-qrem}
\end{equation}
with $y_{\typ} \geq 0$ necessarily required. Clearly, however, this condition cannot be satisfied, since
for any finite $\mathcal{O}(1)$ values of $\Gamma$ and $W_\tot$, the denominator in Eq.~\eqref{eq:ytyp-qrem} 
becomes negative for $N\gg1$.  A self-consistent solution for $y_\typ$  is thus not possible. This shows that a many-body localised phase in the middle of the band can never be stabilised for the QREM, and the model is always delocalised, in accordance with previous results~\cite{laumann2014many,baldwin2016manybody}.

We emphasise that the qualitatively different results arising in the two opposite limiting cases considered 
in the two sub-sections, have been obtained from the same basic calculation; 
all that differed was the precise form of the covariance matrix, $\mathbf{C}_\mathbf{x}$, which encodes the 
correlations between Fock-space site energies. Such a general  theory, which predicts the possibility or otherwise of a many-body localised phase based on minimal properties of the Hamiltonian on the Fock-space, 
independently of microscopic details, is one of the main results of this work.

\subsubsection{Arbitrary correlations 
\label{subsubsec:arbitcor}}

We now move away from the two limits of $\rho_r= 1$ (for finite $r$) and $\rho_r= 0$,  
to discuss the case $\rho_{r}$ strictly less than 1. For analytic convenience we continue to consider $\rho_2=\rho_1^2$
(though without loss of generality, as noted in Sec.\ \ref{subsection:selfcontheory}).

The basic expression Eq.~\eqref{eq:Fk} for the Fourier transform of the distribution of $y=\Delta/\eta$, 
can be written as
\begin{equation}
F_k^{\pd}(k) = \int dx_{I_0}^{\pd}\, \mathcal{P}_1^{\pd}(x_{I_0}^{\pd})\left[f(k,x_{I_0}^{\pd})\right]^N
\label{eq:fkNe}
\end{equation}
with
\begin{equation}
f(k,x_{I_0}^{\pd}) = \int dx_{I_\alpha}^{\pd}\,\tilde{\mathcal{P}}_1^{\pd}(x_{I_\alpha}^{\pd}\vert x_{I_0}^{\pd})
\exp\Big(\frac{ik\kappa}{\mu_\mathcal{E}^2x_{I_\alpha}^2}\Big),
\label{eq:fk}
\end{equation}
where the conditional distribution $\tilde{\mathcal{P}}_1(x_{I_\alpha}\vert x_{I_0})$ is $N$-independent.
While the full form of the integral in Eq.~\eqref{eq:fk} is rather cumbersome, the essential 
point is that one can express $[f(k,x_{I_0})]^N$ as
\begin{equation}
\begin{split}
&[f(k,x_{I_0}^{\pd})]^N =
\\
&\exp\left[N\ln\left(1+\sqrt{\frac{|k|\kappa}{W_{\tot}^{2}N}}\,Q\left(\sqrt{\frac{|k|\kappa}{W_{\tot}^{2}N}};\rho_1^{\pd}\right)\right)\right],
\end{split}
\label{eq:fkN}
\end{equation}
where the function $Q$ encodes the details of the integral and all $N$-dependence is explicit.
In the relevant limit $N\gg1$, the leading contribution to Eq.~\eqref{eq:fkN} is clearly
\begin{equation}
[f(k,x_{I_0}^{\pd})]^N \overset{N\gg 1}{\sim}\exp\left[\sqrt{|k| \tilde{\kappa}}\,Q\left(0;\rho_1^{\pd}\right)\right],
\label{eq:fkN-k0}
\end{equation}
where for brevity we have defined
\begin{equation}
\tilde{\kappa} ~:=~ \frac{N\kappa}{W_{\tot}^{2}} ~\propto ~ N .
\end{equation}
From Eqs.\ \eqref{eq:fkN-k0} and \eqref{eq:fkNe} $F_k(k)$ is thus a function solely of $k\tilde{\kappa}$,
$F_k(k) \equiv\hat{g}(k\tilde{\kappa})$, with all $N$-dependence encoded in $\tilde{\kappa}$.
This in turn implies that its Fourier transform, $F_{y}(y,y_{\typ})$, is of form
\begin{equation}
F_y^{\pd}(y,y_\typ^{\pd}) = \frac{1}{\tilde{\kappa}}\,g\left(\frac{y}{\tilde{\kappa}}\right)
\label{eq:Fy-scaling}
\end{equation}
with $g(y)$ the inverse transform of $\hat{g}(k)$. The self-consistency condition 
$\ln y_{\typ} = \int dy \ln(y) F_{y}(y,y_{\typ})$ thus becomes
\begin{equation}
\ln y_\typ^{\pd} = \ln \left(\frac{N\kappa}{W_{\tot}^{2}}\right) + \underbrace{\int dv\, g(v)\,\ln v}_{V}
\end{equation}
where, importantly, $V$ does not depend on $N$ (and reduces to $V=\gamma+\ln 2$ in the QREM limit $\rho_{2}=0=\rho_{1}$).
Since $\kappa =\Gamma^{2}(1+y_{\typ})$ (Eq.\ \eqref{eq:yI0-strong}), this yields
\begin{equation}
y_\typ^{\pd} = \frac{\Gamma^2 N e^V}{W_{\tot}^2-\Gamma^2 N e^V} ~,
\label{eq:ytyp-scaling-general}
\end{equation}
where all the $N$-dependence is explicit.

The $N$-dependence of Eq.~\eqref{eq:ytyp-scaling-general} has precisely the same form as that for the QREM
(Eq.\ \eqref{eq:ytyp-qrem}). Hence, by the same reasoning as given in Sec.\ \ref{subsubsec:QREMlimit} (under Eq.\ \eqref{eq:ytyp-qrem}), it follows that a self-consistent solution for $y_{\typ}$ is not possible. One thus concludes that
a many-body localised phase cannot be self-consistently stabilised in this general case of $\rho_2 <1$, 
just as found for the QREM limit $\rho_{2}=0$.


\subsection{{Localisation criterion for correlations}
\label{subsubsec:criterion}}

The results presented above allow us to make a general prediction for what conditions the correlations must satisfy 
in order that a stable many-body localised phase is possible. Before stating that result, let us briefly recapitulate the main points arising from this section.

A probabilistic, self-consistent mean-field theory for the imaginary part of the local self-energy on the Fock space can be set up  along the lines of Refs.~\cite{logan2019many,roy2019self}, where an essential step is that many-body energy scales are rescaled with $\sqrt{N}$. The satisfaction, or otherwise, of a self-consistency condition on the rescaled imaginary part of the self-energy signals the presence, or absence, of a stable localised phase. 
At the renormalised second-order level considered, the theory requires as input the $(N+1)$-dimensional joint distribution of the site energies of a Fock-space site and its neighbours, which encodes correlations between the Fock-space site energies. The correlations are thus those at Hamming distance two and one, embodied respectively by 
the covariances $\rho_2$ and $\rho_1$. 
Importantly, however, the relevant distribution $F_y$ of the imaginary part of the self-energy depends solely on $\rho_2$ 
and is independent of $\rho_1$. We exploit this feature  by considering the choice $\rho_1^2=\rho_2$, which provides us with a convenient device to obtain $F_y$ analytically.

With this, we find that in the maximally correlated limit a stable localised phase does indeed exist.
Short- and long-ranged disordered quantum systems with local interactions, studied previously~\cite{logan2019many,roy2019self}, belong to this category. Remarkably, for any deviation away from this limit, localisation does not persist even if the energies are partially correlated. The QREM, which corresponds to the extreme limit of no correlations and falls in this category~\cite{laumann2014many,baldwin2016manybody}, is also captured correctly by the theory.

The results obtained in this section present us with an interesting picture which implies that a localised phase is stable only when the Fock-space site energies are \emph{maximally correlated} at Hamming distance two, $\rho_2=1$. 
For physical models, it is a reasonable requirement to have $\rho_r$ as a smooth and monotonically decreasing function of $r$, at least for finite $r$ such that $r/N \to 0$ as $N\to \infty$.
Under this requirement, $\rho_2=1$ implies that $\rho_r=1$ for all finite $r$. Hence, a central result of this section can be stated as: \emph{a many-body localised phase can be self-consistently stabilised only if the Fock-space site energies are maximally correlated at finite Hamming distances.}  It is not enough for them to be partially correlated for localisation to persist, since any  degree of independence to them, which takes them away from the $\rho_r\to 1$ limit, completely destabilises the localised phase in the thermodynamic limit, and makes the model equivalent to the QREM 
(see also Fig.\ \ref{fig:schematic-pd}).

The core predictions of this section will also be tested in Sec.\ \ref{sec:microscopic} by exact diagonalisation, using well-established numerical diagnostics.


\section{Microscopic models 
\label{sec:microscopic}}

Having obtained a criterion that correlations of the Fock-space site energies must satisfy for a stable many-body localised phase to occur, we now turn our attention to microscopic model Hamiltonians. 

Three types of models are considered: a disordered nearest-neighbour quantum Ising chain (Sec.\ \ref{subsection:NNTFI}),
disordered quantum $p$-spin models at finite $p$ (Sec.\ \ref{subsection:pspinmods}),  
and a quantum random energy model modified such that correlations in its  Fock-space site energies decay exponentially with Hamming distance  (Sec.\ \ref{subsection:expREM}).
We show that the distributions of  Fock-space site energies over both disorder realisations and Fock space,
$\mathcal{P}_\mathcal{D}$, are multivariate Gaussians; rendering the theory above applicable to them, and
enabling us to predict whether or not a particular model hosts a localisation transition.

For each model, the predictions of the self-consistent theory for the stability or otherwise of a localised  phase
are then tested numerically, by studying the putative many-body localisation transition (or its absence) using exact diagonalisation. The numerical and theoretical results are indeed found to be in mutual  agreement.

We study in particular the statistics of level-spacing ratios, participation entropies of eigenstates on the Fock space, and bipartite entanglement entropies. These quantities are complementary, and ubiquitously studied in the context of many-body localisation; and in our case, reassuringly, they concurrently predict the presence or absence of a localisation transition. Before delving into the models, we describe briefly these numerical diagnostics and their behaviour in the two phases.

The level-spacing ratio, $r_n$,\footnote{Here, to be consistent with earlier literature, we use the same symbol ($r$) as for Hamming distance.  The meaning of $r$ will however be clear from context.}  is defined as~\cite{oganesyan2007localisation,atas2013distribution}
\begin{equation}
r_n = \frac{\min(s_n,s_{n+1})}{\max(s_n,s_{n+1})} ~~~~: s_n=E_n-E_{n-1},
\label{eq:levelstats}
\end{equation}
where the $E_n$ denote consecutive eigenvalues of the Hamiltonian.  Signifying an ergodic phase, the presence of level repulsion  leads to $r_n$ following a Wigner-Dyson distribution with the appropriate symmetries (in our case the Gaussian Orthogonal Ensemble, GOE). In the localised phase by contrast the eigenvalues are 
uncorrelated, leading to a Poisson distribution for $r_n$. The former corresponds to $\braket{r}_\mathrm{GOE}\simeq 0.536$
and the latter to $\braket{r}_\mathrm{Poisson}=2\ln2-1\simeq 0.386$~\cite{atas2013distribution}.
For finite systems, $\braket{r}$ drifts with increasing system size towards $\braket{r}_\mathrm{GOE}$ in a delocalised phase, whereas in the localised phase it drifts towards $\braket{r}_\mathrm{Poisson}$.
A critical point is thus indicated by the crossing of the data for various system sizes.

The first participation entropy for an eigenstate $\ket{\psi}$, defined as 
\begin{equation}
S^\mathrm{PE}_1(\ket{\psi})=-\sum_{I}\vert\braket{\psi\vert I}\vert^{2}\log\vert\braket{\psi\vert I}\vert^{2},
\label{eq:participation}
\end{equation}
is a measure of the support of the eigenstate on the Fock space, with its dependence on  the Fock-space dimension given by
\begin{equation}
S^\mathrm{PE}_1(\ket{\psi}) = \alpha_1\ln N_\mathcal{H} +\mathcal{O}(\ln\ln N_\mathcal{H}).
\label{eq:PEscaling}
\end{equation}
In the delocalised phase an eigenstate has support on a finite fraction of Fock space, whence $\alpha_1=1$. 
In a localised phase by contrast, eigenstates have support on a vanishing fraction of the Fock-space sites,  resulting in $\alpha_1<1$~\cite{deluca2013ergodicity,luitz2015many,mace2019multifractal}. The departure of $\alpha_1$ from unity thus signals a localisation transition.

Finally, the bipartite entanglement entropy $S^\mathrm{EE}$  also acts a diagnostic for the two phases, exhibiting a volume law in a delocalised phase but an area law in a localised phase~\cite{kjall2014many,luitz2015many,luitz2016long,bauer2013area,lim2016many}. For a system of $N$ spins-1/2, we refer to the subsystem comprising spins from 1 through $N/2$ as $A$, and the remainder as $B$. The entanglement entropy between subsystems $A$ and $B$ in the eigenstate 
$\ket{\psi}$ is then 
\begin{equation}
S^\mathrm{EE} = -\mathrm{Tr}_A^{\pd}[\rho_A^{\pd}\ln\rho_A^{\pd}]\sim\begin{cases}N~~:~ \mathrm{delocalised}\\ N^0 ~:~ \mathrm{localised}\end{cases}
\label{eq:entanglement}
\end{equation}
with $\rho_A=\mathrm{Tr}_B\ket{\psi}\bra{\psi}$ the reduced density matrix of subsystem $A$.

In our numerical results  we focus on the centre of the spectrum, considering 25-50 eigenstates 
(depending on system size) with their eigenvalues closest to $\mathrm{Tr}[H]$.
We also add that specific values of critical disorder strengths should not be regarded
as numerically definitive; our purpose is to ascertain the presence or absence of a transition, rather 
than to locate it particularly accurately.


\subsection{Nearest-neighbour interacting disordered spin-1/2 chain}
\label{subsection:NNTFI}

The disordered transverse-field Ising chain is a well-established model for many-body localisation~\cite{kjall2014many,imbrie2016many,roy2018exact}, 
\begin{equation}
H_\mathrm{TFI} = \sum_{\ell=1}^N [J_\ell^{\pd}\sigma^z_\ell\sigma^z_{\ell+1} + h_\ell^{\pd}\sigma^z_\ell] +\Gamma\sum_{\ell=1}^N\sigma^x_\ell
\label{eq:Htfi}
\end{equation}
(with periodic boundary conditions). The independent random fields $h_{\ell}$ are chosen from a normal distribution with zero mean and a standard deviation $W_h$, i.e.\ $h_l\sim \mathcal{N}(0,W_h^2)$. 
We also consider the random couplings $J_{\ell}$ to be normally distributed, with mean $J$ and standard deviation 
$W_J$; $J_l\sim\mathcal{N}(J,W_J^2)$. The random couplings thus separate into a constant part $J$ and a disordered 
part $J_\ell^\prime$, where the latter is normally distributed with zero mean and standard deviation $W_J$.
As in Eqs.~\eqref{eq:mvg-eiprime} and \eqref{eq:mvg-eizero}, the Fock-space site energies hence
separate into a disorder independent and a disorder dependent part, given respectively by
\begin{equation}
\begin{aligned}
\mathcal{E}_I^0&=J\sum^{N}_{\ell=1}s^{(I)}_{\ell}s^{(I)}_{\ell+1},\\
\mathcal{E}_I^\prime&=\sum^{N}_{\ell=1}\big[J_\ell^\prime ~s^{(I)}_{\ell}s^{(I)}_{\ell+1}
+h_{\ell}^{\pd}s^{(I)}_{\ell}\big].\\
\end{aligned}
\label{eq:decompose-fs-energy-nn}
\end{equation}


\subsubsection{Distribution and correlations of Fock-space site energies}
\label{subsubsection:NN-TFI-dists}

\begin{figure}
\includegraphics[width=\columnwidth]{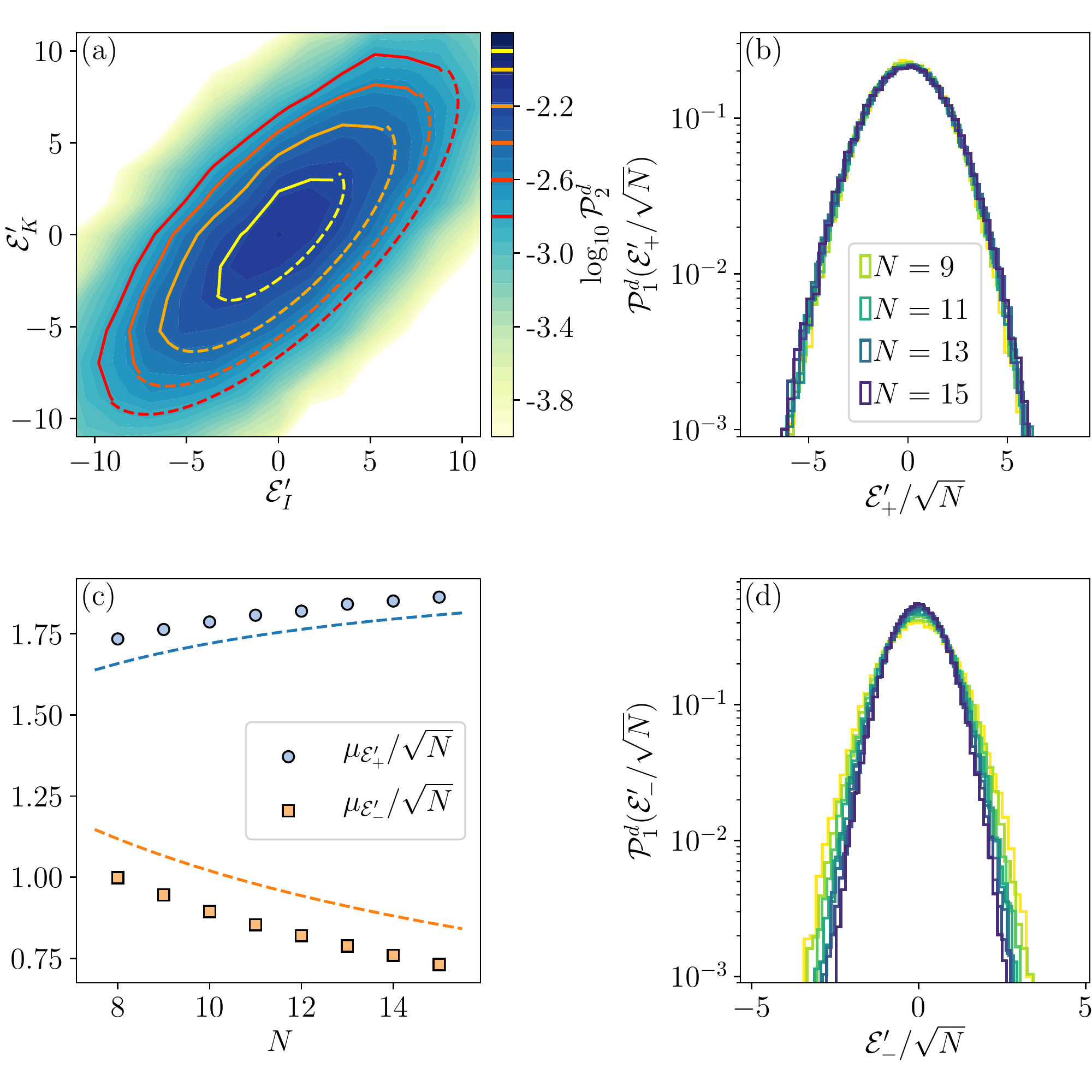}
\caption{
Results for the bivariate distribution  of Fock-space site energies over disorder realisations, for the disordered Ising chain Eq.~\eqref{eq:Htfi} with $W_J=1=W_h$. (a) The distribution is shown as a colour-map. Representative contours are highlighted, with the solid line denoting numerical results and the dashed lines showing the contours corresponding to a bivariate Gaussian with the same moments. Panels (b) and (d) show the univariate distribution of $\E_\pm^\prime$ for different system sizes, $N$, appropriately rescaled. The standard deviations of these distributions, 
$\mu_{\E_\pm^\prime}$, are shown in (c) where the dashed lines correspond to the analytic predictions, 
$\mu_{\E_\pm^\prime}^{2}=C_d(0)\pm C_d(r)$ with $C_d(r)$ given by Eq.~\eqref{eq:tfi-Cdr}. Distributions are generated over $5\times 10^4$ disorder realisations for all $N$; for panel (a), $N=15$.}
\label{fig:pd-nn}
\end{figure}

We do not have a general analytical demonstration of the Gaussian nature of $\P_\mathcal{D}$ in the
thermodynamic limit (though this can be given for the one-body distribution $\P_1(\E_{I})$~\cite{welsh2018simple}).
We thus study the problem numerically in the first instance, focussing on the bivariate case $\mathcal{D}=2$ for
convenience with visual representation of the results; and considering in turn
the distribution $\P_2^d$ over disorder realisations, followed by that over the Fock space, $\P_2^F$.

To proceed, we first numerically generate  
$\mathcal{P}_2^{d}(\mathcal{E}_I^\prime,\mathcal{E}_K^\prime)$ with $r_{IK}=r$. 
The two univariate distributions of $\E_\pm^\prime =(\E_I^\prime\pm\E_K^\prime)/\sqrt{2}$ are then studied. 
If $\mathcal{P}_2^{d}$ is a bivariate Gaussian, with $C_d(0)$ and $C_d(r)$ as the diagonal and offdiagonal elements of its covariance matrix, then each of $\E_\pm^\prime$ should also be normally distributed with variances 
$\mu^{2}_{\mathcal{E}^{\prime}_{\pm}} = C_d(0)\pm C_d(r)$. We check for this by comparing the numerical distributions to Gaussian distributions with the same covariances. Such an analysis is useful, because it not only probes the Gaussian nature of the distribution but also tests the correctness of the correlations thereby obtained. The same analysis is employed in considering $\mathcal{P}^F$.

To generate the distribution $\P_2^d(\E_I^\prime,\E^\prime_K)$ we consider without loss of generality the pair of
Fock-space sites $\ket{I}=\ket{\up\cdots\up\up\up}$ and $\ket{K}=\ket{\up\cdots\up\dn\dn}$, separated by a Hamming distance $r_{IK}=2$. The numerically generated bivariate distribution is shown as a colour-map in 
Fig.~\ref{fig:pd-nn}(a). The elliptical contours seen are already indicative of the Gaussian nature of the distribution. 
The distribution is also symmetric about the $\E_I^\prime=\E_K^\prime$ line. Hence, a direct comparison between the numerically generated distribution and the bivariate Gaussian with the same moments, can be obtained by plotting 
contours corresponding to the numerically generated distribution in the region $\E_I^\prime<\E_K^\prime$, and those corresponding to the bivariate Gaussian in the $\E_I^\prime>\E_K^\prime$ region. Such an exercise, shown in Fig.~\ref{fig:pd-nn}(a), leads to a rather symmetric picture, confirming the Gaussian nature of the distribution.

 Next, we study the distributions of $\E^\prime_\pm = (\E_I^\prime\pm\E_K^\prime)/\sqrt{2}$ in panels (b) and (d) respectively. Consistent with $\P_2^d$ being a bivariate Gaussian with a covariance matrix of form $\left(\begin{smallmatrix}C(0) &&C(r)\\C(r) &&C(0)\end{smallmatrix}\right)$, $\P_1^d(\E_\pm^\prime)$ are univariate Gaussians, as indicated by the inverted parabolic shape on log-linear axes.  The distribution of $\E_+^\prime/\sqrt{N}$ (panel (b)) shows evident signs of converging in the thermodynamic limit to a normal distribution with a finite standard deviation; indeed the $N$-dependence of its standard deviation $\mu_{\E^\prime_+}/\sqrt{N}$ (Fig.~\ref{fig:pd-nn}(c))
shows qualitative agreement with the analytic prediction of it (Eq.\ \eqref{eq:tfi-Cdr} below), which 
in the thermodynamic limit tends to $\sqrt{2(W_{J}^{2}+W_{h}^{2})}$ ($=2$ for the parameters considered).
The distribution of $\E_-^\prime/\sqrt{N}$ by contrast (Fig.~\ref{fig:pd-nn}(d)) narrows with increasing $N$. 
The  $N$-dependence of its standard deviation $\mu_{\E^\prime_-}/\sqrt{N}$ (Fig.~\ref{fig:pd-nn}(c)) likewise shows qualitative agreement with the analytic  prediction arising from Eq.\ \eqref{eq:tfi-Cdr}, which for any finite $r$ predicts $\mu_{\E^\prime_-}/\sqrt{N}$ to vanish $\propto 1/\sqrt{N}$ as the thermodynamic limit is approached. This of course corresponds to  $(\E_I^\prime-\E_K^\prime)/\sqrt{N}\to 0$ in the thermodynamic limit, as known for short-ranged models which fall in the maximally correlated class~\cite{logan2019many,roy2019self}.

\begin{figure}
\includegraphics[width=\columnwidth]{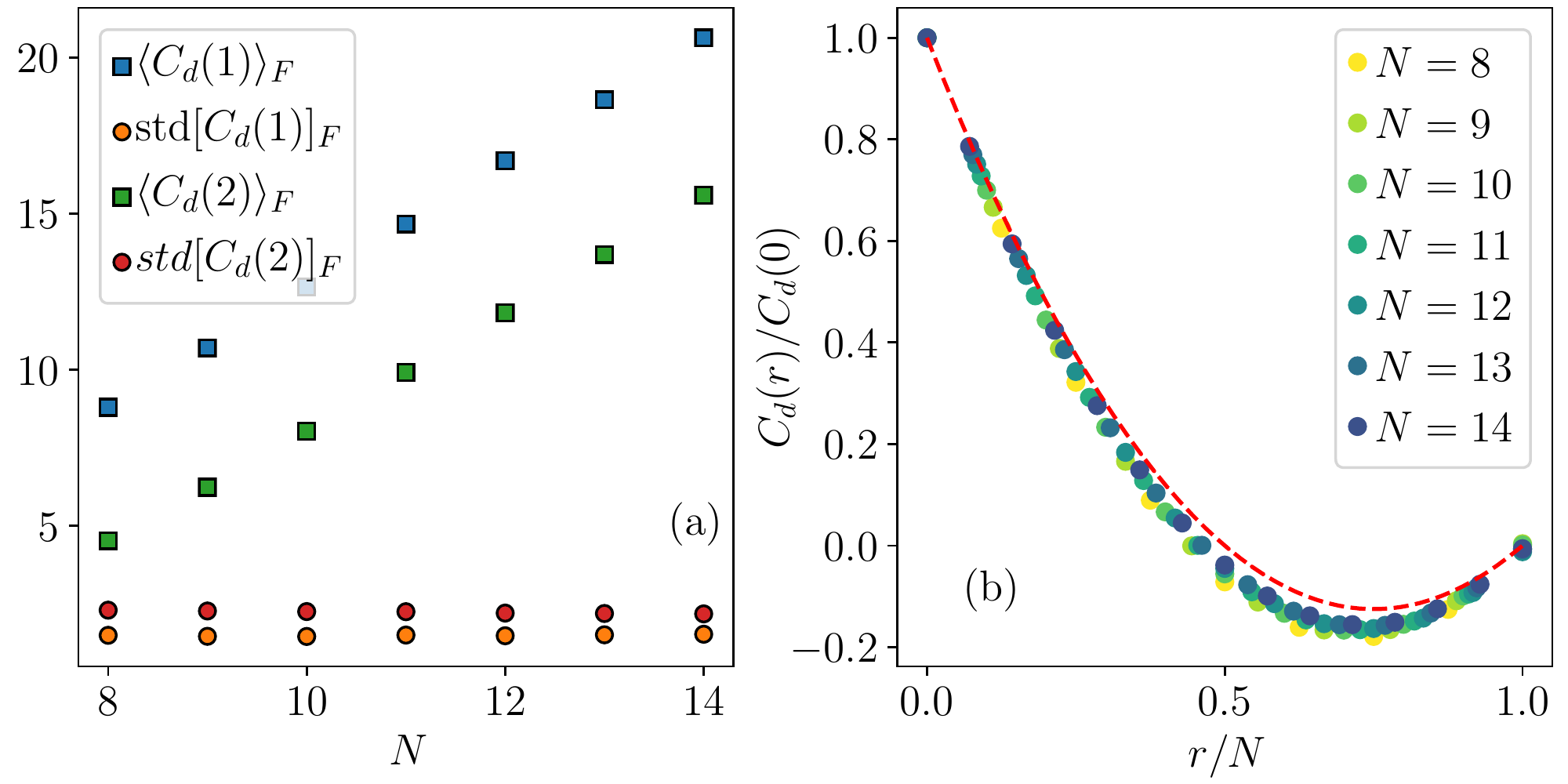}
\caption{(a) The mean and standard deviation of $C_d^{IK}(r)$ over all Fock-space site pairs $(I,K)$ such that 
$r_{IK}=r$, as a function of $N$ for $r=1$ and 2. The linear  $N$-dependence of the former and the constant behaviour of the latter implies that in the thermodynamic limit $\tilde{C}_d^{IK}(r)=C_d^{IK}(r)/N$ depends only on $r$. (b) The Fock-space averaged $C_d(r)/C_d(0)$ as a function of $r/N$ collapses for different $N$, implying that it is indeed a function of $r/N$ alone. The red dashed line corresponds to the analytic prediction from  Eq.~\eqref{eq:tfi-Cdr}. For the plots, 
$W_J=1=W_h$.}
\label{fig:covmat-nn}
\end{figure}

While the above analysis shows that $\P_2^d(\E_I^\prime,\E_K^\prime)$ is a bivariate Gaussian, we still need to demonstrate that the covariance matrix characterising the distribution depends only on the Hamming distance and not the specific $I$ and $K$ chosen. To show this numerically, we obtain $C_d^{IK}(r)$ by averaging over disorder for all pairs of Fock-space sites $(I,K)$, which satisfy $r_{IK}=r$.
A distribution of $\{C_d^{IK}(r)\}$ can then be generated over the Fock-space site pairs. Since $C_d^{IK}(r)\propto N$, it is in fact $\tilde{C}_d^{IK}(r)=C_d^{IK}(r)/N$ which admits a well-defined thermodynamic limit.  As shown in Fig.~\ref{fig:covmat-nn}(a), we find numerically that $\tilde{C}_d^{IK}(r)$ has a finite, $N$-independent mean, whereas its standard deviation decays systematically with $N$.
This implies that in the thermodynamic limit, $C_d$ is indeed independent of the specific Fock-space sites and depends  only on the Hamming distance. In fact, as shown in Fig.~\ref{fig:covmat-nn}(b), the numerically calculated $C_d^{IK}(r)$, when averaged over the Fock-space site pairs (as in Eq.\ \eqref{eq:covariance-def}), shows excellent agreement with the analytical prediction of  Eq.~\eqref{eq:tfi-Cdr} below. We also add that a finite $\mathrm{std}[C_d(r)]$ in a finite-sized system is the origin of the quantitative discrepancy between the numerical and analytical results for $\mu_{\E_\pm^\prime}/\sqrt{N}$ in Fig.~\ref{fig:pd-nn}(c), since the numerical results there correspond to a particular pair $I$ and $K$.

\begin{figure}
\includegraphics[width=0.49\columnwidth]{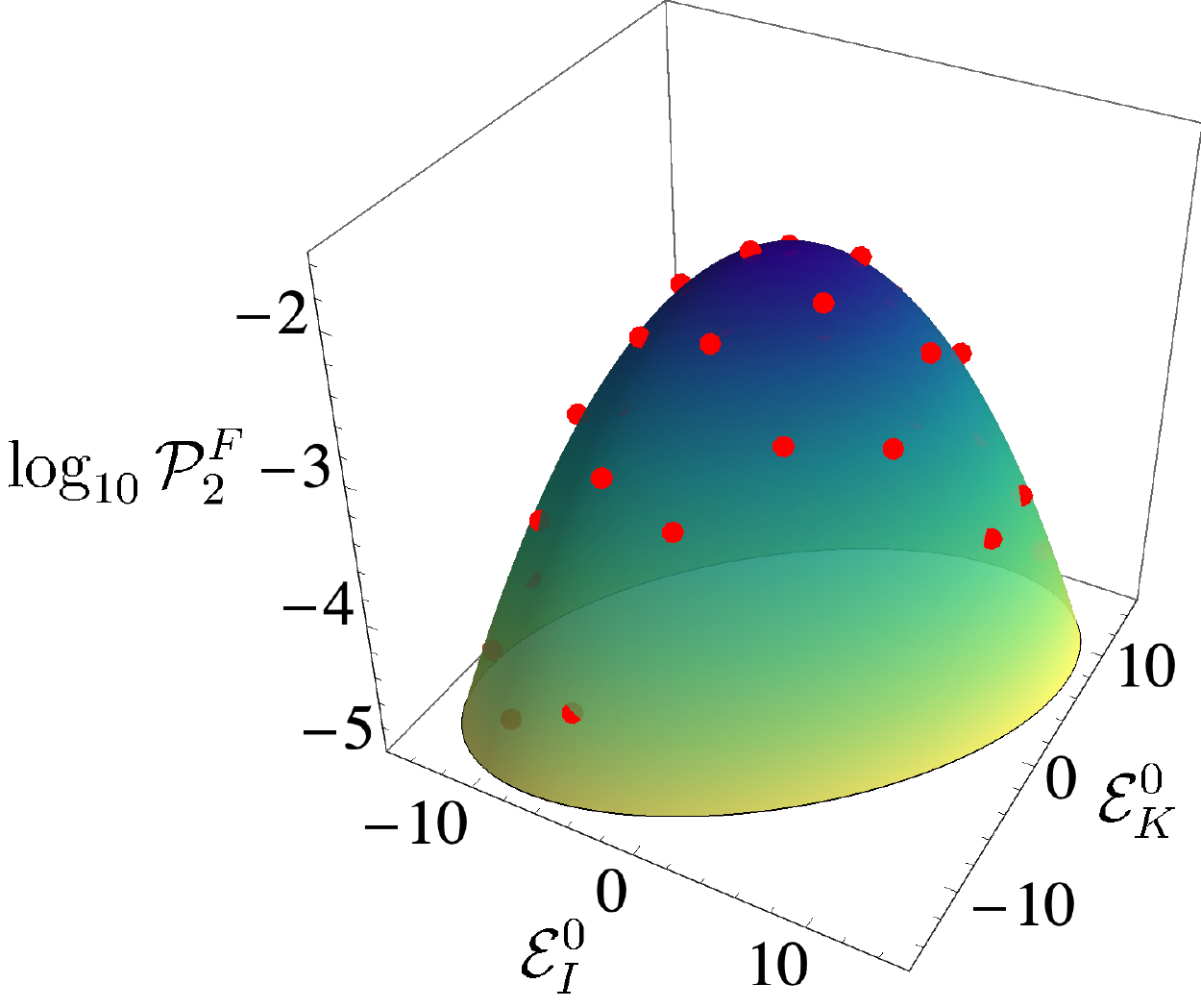}
\includegraphics[width=0.49\columnwidth]{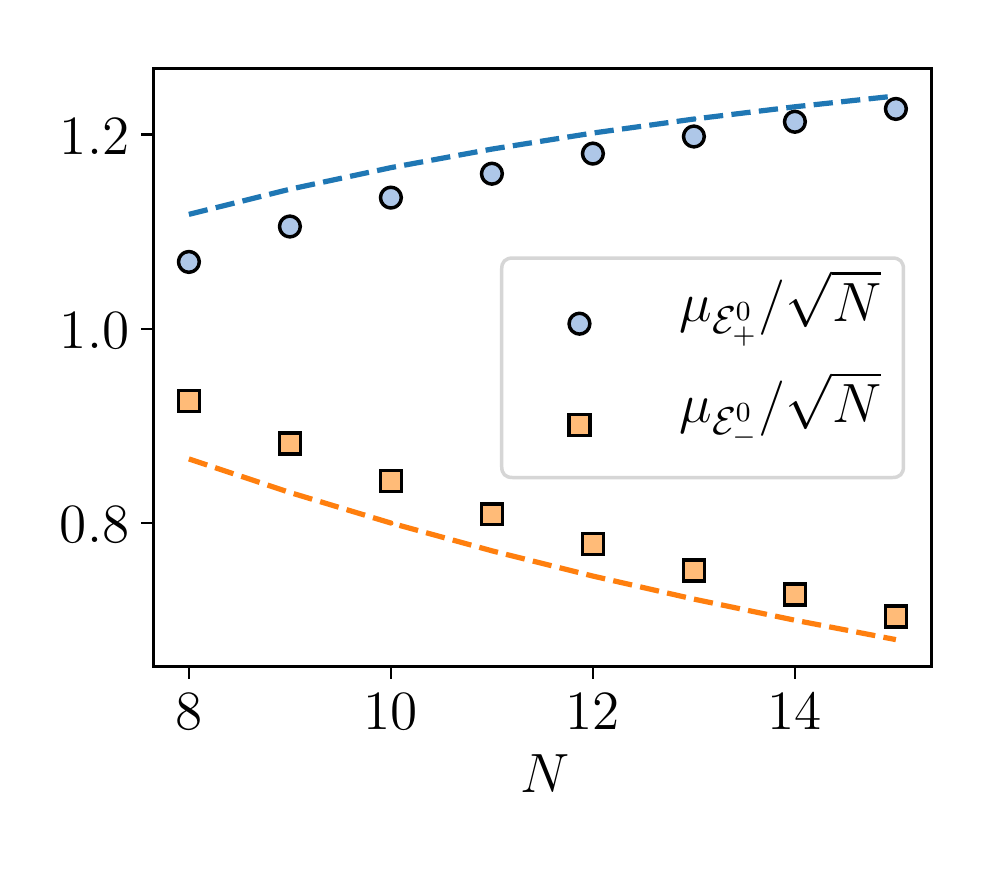}
\caption{Results for $\P^F_{2}$ for the disordered Ising chain, with $J=1$. \emph{Left}: Red dots represent the numerically generated histogram of the tuple $(\E_I^0,\E_K^0)$ over all Fock-space sites such that $r_{IK}=2$ 
(for $N=15$). The underlying surface is a bivariate Gaussian with the moments calculated from the numerical histogram. \emph{Right}: The standard deviations of the distributions of $\E_\pm^0$ rescaled with $\sqrt{N}$,  showing good agreement with the analytic predictions of Eq.~\eqref{eq:tfi-cfr} (dashed lines).}
\label{fig:pf-nn}
\end{figure}

We turn now to the corresponding distribution over the Fock space, $\P_2^F$. For this Ising chain, 
$\mathcal{E}_I^0=J\sum_{\ell}\ssh[I]\ell\ssh[I]{\ell+1}$. For a finite-system of size $N$, however,
$\mathcal{E}_I^0$ can only take integer values, $N,N-2,N-4,\cdots,-N$, and hence the spectrum of $\E_I^0$ is strongly degenerate. This naturally renders difficult a smooth sampling of the distribution and its representation as a colour-map, for the $N$-range numerically accessible in practice. We can however generate a two-dimensional histogram of the set of energies $(\E_I^0,\E_K^0)$ over all Fock-space sites such that $r_{IK}=r$, and overlay the histogram with a bivariate Gaussian of the same moments. As seen in Fig.~\ref{fig:pf-nn}, the two seem to be in excellent agreement given that the data was generated with $N=15$. 

In direct parallel to the analysis for $\P_2^d$, we also study numerically the distributions of 
$\E_\pm^0=(\E_I^0\pm\E_K^0)/\sqrt{2}$, and plot in Fig.~\ref{fig:pf-nn} the $N$-dependence of their rescaled standard deviations, $\mu_{\mathcal{E}^{0}_{\pm}}/\sqrt{N} = [C_F(0)\pm C_F(r)]^{1/2}/\sqrt{N}$.
Corresponding analytical results can again be obtained, from Eq.~\eqref{eq:tfi-cfr} below for $C_{F}(r)$.
As seen in Fig.~\ref{fig:pf-nn}, the agreement between numerics and the analytical expressions (dashed lines) 
is very good, becoming increasingly so with increasing $N$.  The analytical results again show that 
$\mu_{\mathcal{E}^{0}_{+}}/\sqrt{N}$ approaches a finite value in the thermodynamic limit, while  
$\mu_{\mathcal{E}^{0}_{-}}/\sqrt{N}$ correspondingly vanishes,  symptomatic of $(\E_I^0-\E_K^0)/\sqrt{N}\to 0$ in the thermodynamic limit. Overall, the numerical analysis is entirely consistent with $\P_2^F$ also being a bivariate Gaussian.


\subsubsection{Analytic results for covariances}
\label{subsubsection:analyticcovs}

We now sketch the derivations of the analytical results for the covariances referred to above; beginning 
with the disorder-dependent part of the Fock-space site energies. Consider two Fock-space sites $I$ and $K$, with a mutual Hamming distance of $r$. Since $\braket{\E_{I}^\prime}=0$ for any $I$, the  matrix element in 
$C_d^{IK}=\braket{\E_{I}^\prime \E_{K}^\prime}$ is 
\begin{equation}
\begin{aligned}
C_d^{IK} &=\sum_{\ell,m}[\braket{J^\prime_\ell J^\prime_m}s^{(I)}_{\ell}s^{(I)}_{\ell+1}s^{(K)}_{m}s^{(K)}_{m+1}
+
\braket{h_\ell^{\pd} h_m^{\pd}}s_\ell^{(I)}s_m^{(K)}]\\
&=\sum_{\ell}[\underbrace{W_J^2(s^{(I)}_{\ell}s^{(K)}_{\ell})(s^{(I)}_{\ell+1}s^{(K)}_{\ell+1})}_{C_{d,J}^{IK}}
+
\underbrace{W_h^2s_\ell^{(I)}s_\ell^{(K)}}_{C_{d,h}^{IK}}]
\end{aligned}
\label{eq:cdIK-tfi}
\end{equation}
(using $\braket{J^\prime_\ell J_m^\prime}=W_J^2\delta_{\ell m}$ and $\braket{h_\ell h_m}=W_h^2\delta_{\ell m}$).

$C_{d,J}^{IK}$ in Eq.~\eqref{eq:cdIK-tfi} can be analysed as follows. Since the Hamming distance between $I$ and $K$ is $r$, there are $r$ real-space sites with $\ssh[I]\ell\ssh[K]\ell =-1$, while $\ssh[I]\ell\ssh[K]\ell =+1$ for the remaining
$N-r$ sites. Suppose that $\ssh[I]m\ssh[K]m =-1$ for a site $m$.
Then on site $m+1$ the probability that $\ssh[I]{m+1}\ssh[K]{m+1} =-1$ is $\phi = (r-1)/(N-1)$, since $r-1$ of the remaining $N-1$ sites have different spins in $I$ and $K$. Naturally, the probability that 
$\ssh[I]{m+1}\ssh[K]{m+1} =+1$ is $1-\phi = (N-r)/(N-1)$.  Likewise, if $\ssh[I]m\ssh[K]m =+1$ for site $m$, 
then the probability that $\ssh[I]{m+1}\ssh[K]{m+1} =-1$ is $r/(N-1)$, and the probability that 
$\ssh[I]{m+1}\ssh[K]{m+1} =+1$ is $(N-r-1)/(N-1)$.
Hence $C_{d,J}^{IK}$ can be expressed as
\begin{equation}
\label{eq:cdIK-tfi-r-holding}
\begin{split}
C_{d,J}^{IK} = & W_J^2\Bigg[-r\left(-\frac{r-1}{N-1} + \frac{N-r}{N-1}\right)
\\
&+ (N-r)\left(\frac{N-r-1}{N-1}- \frac{r}{N-1}\right)\Bigg],
\end{split}
\end{equation}
with the first term corresponding to the $r$ sites with $\ssh[I]\ell\ssh[K]\ell =-1$, and the second to the $N-r$ sites with $\ssh[I]\ell\ssh[K]\ell =+1$. Note that  Eq.~\eqref{eq:cdIK-tfi-r-holding} depends only on the Hamming distance $r$. 
To leading order as $N\to \infty$, it reduces to
\begin{equation}
C_{d,J}(r)=NW_J^2\left(1-2\frac{r}{N}\right)^2.
\label{eq:tfi-cdrJ}
\end{equation}

The second term in Eq.~\eqref{eq:cdIK-tfi}, $C_{d,h}^{IK}$, is simple to analyse. Since there are $r$ sites with
$\ssh[I]\ell\ssh[K]\ell =-1$ and $N-r$ where it is +1, it follows trivially that
\begin{equation}
C_{d,h}(r)=NW_h^2\left(1-2\frac{r}{N}\right)
\label{eq:tfi-cdrh}.
\end{equation}
Eqs.~\eqref{eq:tfi-cdrJ} and \eqref{eq:tfi-cdrh} thus give the result for the disorder-dependent contribution to the covariance matrix, 
\begin{equation}
C_d(r) = NW_J^2\left(1-2\frac{r}{N}\right)^2+NW_h^2\left(1-2\frac{r}{N}\right).
\label{eq:tfi-Cdr}
\end{equation}

\begin{figure*}
\includegraphics[width=\linewidth]{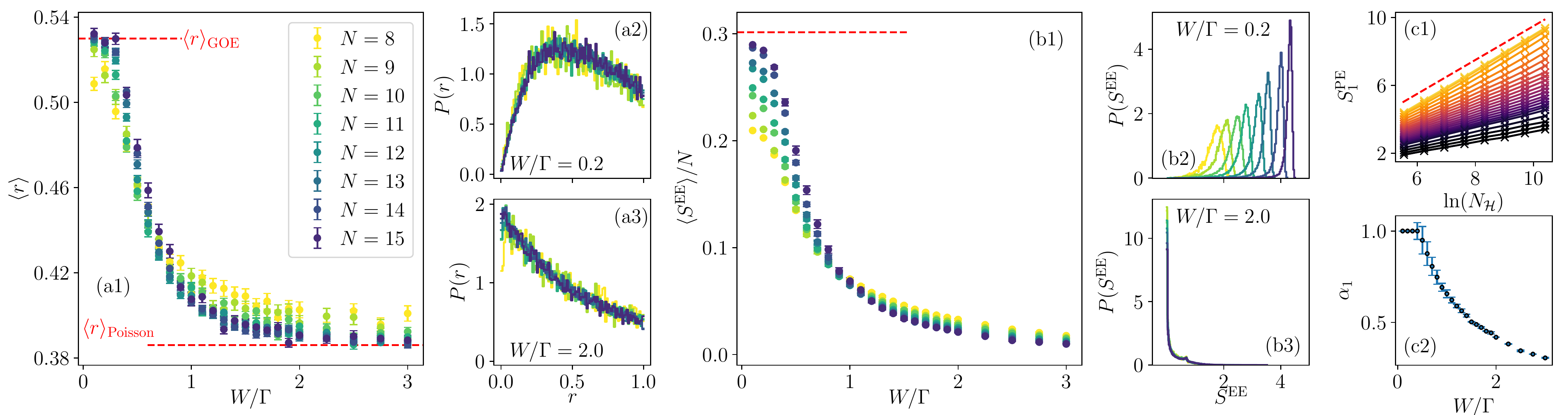}
\caption{Exact diagonalisation results for the disordered quantum Ising model Eq.~\eqref{eq:Htfi}, for $W_J=W_h\equiv W$, $J=0$ (with $\Gamma=1$). Panel (a1): mean level-spacing ratio, $\braket{r}$, as a function of $W/\Gamma$ for different system sizes $N$. At low disorder strengths, $\braket{r}$ approaches the GOE value with increasing $N$, whereas at strong disorder it approaches the Poisson value. Panels (a2) and (a3) show respectively the Wigner-Dyson and Poisson distributions of $r$ in the  delocalised and localised phases. Panel (b1):  bipartite entanglement entropy (with
red dashed line showing the Page value~\cite{page1993average} for an ergodic state). Here again, a crossing of the data for various $N$ indicates a phase transition. Volume and area laws for the entanglement entropy are confirmed by the distributions of $S^\mathrm{EE}$ shown in panels (b2) and (b3), at weak and strong disorder respectively. Panel (c1): first participation entropy, $S^\mathrm{PE}_1$, as a function of $\ln(N_\mathcal{H})$ for various disorder strengths (lighter colours denote weaker disorder strengths). The dashed line shown has unit gradient, for ease of visualisation.  
The linear behaviour is used to extract the slope, $\alpha_1$, shown in (c2), whose deviation from unity indicates a transition to the localised phase. Data were obtained by averaging over 1000 disorder realisations, with statistical errors estimated using a standard bootstrap error analysis.}
\label{fig:ed-nn}
\end{figure*}

Consider now the covariance matrix $C_F(r)$ corresponding to $\P^F_\mathcal{D}$. Note that $\overline{\E^0}=0$ trivially, as $\braket{\ssh[I]\ell\ssh[I]\ell}_{F} =0$.  Hence $C_F(r)\equiv\braket{\E_I^0\E_K^0}_F$, where it is understood that 
the Fock-space average is over all Fock-space site pairs separated by an Hamming distance of $r$. We thus have
\begin{equation}
C_F(r) = J^2\sum_{\ell,m}\braket{s_\ell^{(I)}s_{\ell+1}^{(I)}s_m^{(K)}s_{m+1}^{(K)}}_F,
\end{equation}
where the Fock-space averaging reduces to appropriately contracting spins pairwise, since any isolated spin would average to zero over the Fock space. The only non-trivial contribution to the sum is then given by the term 
with $\ell=m$ such that  $C_F(r)=J^2\sum_\ell\braket{[\ssh[I]\ell\ssh[K]\ell][\ssh[I]{\ell+1}\ssh[K]{\ell+1}]}_F$. 
Note that this form is identical to that of $C_{d,J}^{IK}$ in Eq.~\eqref{eq:cdIK-tfi}, 
so precisely the same arguments as given after Eq.~\eqref{eq:cdIK-tfi} follow.  Hence one immediately concludes that 
\begin{equation}
C_F(r)=NJ^2\left(1-2\frac{r}{N}\right)^2.
\label{eq:tfi-cfr}
\end{equation}

Since both $\P_\mathcal{D}^d$ and $\P_\mathcal{D}^F$ are multivariate Gaussians, the distribution over both disorder and Fock-space, $\P_\mathcal{D}=\P_\mathcal{D}^d\ast\P_\mathcal{D}^F$, is also a Gaussian with a covariance matrix which is simply a sum of $C_d$ and $C_F$. From Eqs.~\eqref{eq:tfi-Cdr} and \eqref{eq:tfi-cfr}, we thus have
\begin{equation}
C(r)=N\left[(J^2+W_J^2)\left(1-2\frac{r}{N}\right)^2 + W_h^2\left(1-2\frac{r}{N}\right)\right],
\label{eq:tfi-cr}
\end{equation}
from which $\rho_{r} =C(r)/C(0)$ follows. For any finite $r$ in the limit of $N\to\infty$, it follows that $\rho_r\to 1$.
This implies that the model falls in the class of maximally correlated models discussed in Sec.~\ref{subsec:maximally}, and thus should host a many-body localised phase. This is corroborated using exact diagonalisation in the following section. 

We also remark that, from Eq.~\eqref{eq:tfi-cr} for $C(r=0)=\mu_{\E}^{2}$, one can read off directly 
$W_{\tot} =\mu_{\E}/\sqrt{N}$ (Eq.\ \ref{eq:WFSdef}) as $W_\tot =\sqrt{J^2+W_J^2+W_h^2}$; the contribution of $J$ to which reflects the underlying configurational disorder on the Fock space~\cite{logan2019many,roy2019self}.


\subsubsection{Exact diagonalisation results}
\label{subsubsection:ED-NNTFI}

Exact diagonalisation results are shown in Fig.~\ref{fig:ed-nn} for the three numerical diagnostics: level-spacing ratios, entanglement entropies and participation entropies, panels (a)-(c) respectively.
The data are for $W_J=W_h\equiv W$ and $J=0$.

The mean level-spacing ratio as a function of $W/\Gamma$ (panel a1) is seen to cross over with increasing $W/\Gamma$
from the GOE value to that corresponding to a Poisson distribution, with the data for various system sizes crossing 
as the critical disorder strength is traversed. Further evidence for the existence of the two phases is shown in panels 
(a2) and (a3) where, for weak and strong disorder respectively, the distribution of the level-spacing ratio clearly 
follows a Wigner-Dyson and Poisson distribution.

Concomitantly, the mean bipartite entanglement entropy scaled with system size, 
$\langle S^{\mathrm{EE}}\rangle/N$ (with $S^{\mathrm{EE}}$ defined in Eq.~\eqref{eq:entanglement}), 
also shows the signatures of the transition (panel (b1)). 
For low $W/\Gamma$ it drifts towards the Page value~\cite{page1993average} taken by a random ergodic state 
in the Fock space, and hence exhibits a volume-law behaviour. By contrast, at higher values of $W/\Gamma$ it 
decays systematically, indicating an area-law behaviour. This is further clarified in panels (b2) and (b3), where the distributions $P(S^{\mathrm{EE}})$ of the the entanglement entropy over eigenstates and disorder realisations are shown, for disorder strengths in the delocalised phase (panel (b2)) and localised phase ((b3)). 
For low $W/\Gamma$, it is sharply peaked at a value which increases linearly with system size, corresponding to
volume-law behaviour;  while at high $W/\Gamma$ it is peaked at a value which is independent of $N$, confirming the area-law behaviour.

Finally, as evident  in panel (c1), the participation entropy $S_1^\mathrm{PE}$ (defined in Eq.~\eqref{eq:participation}) shows a linear behaviour with $\ln(N_\mathcal{H})$. Its slope, $\alpha_1$, is pinned at 1 for a range of $W/\Gamma$, indicating the delocalised phase. However, at a threshold value of $W/\Gamma$, $\alpha_1$ deviates from unity on entering the localised phase, and thereafter decreases monotonically with increasing $W/\Gamma$ (panel (c2)). 

Each of the three measures thus consistently predict a stable many-body localised phase beyond a threshold $W/\Gamma$, which is entirely in harmony with the predictions from the self-consistent theory.

We also add here that there is a tension between our predictions and the numerical results of 
Ref.~\cite{ghosh2019manybody}, where it was reported that a many-body localised phase 
can exist only if the reduced covariance $\rho_{r}$ has a linear variation in $r/N$. 
However,  the disordered transverse-field Ising model considered in this section provides 
a case in contrast to this claim; for the model is well known (on analytical grounds~\cite{imbrie2016many}) 
to host a many-body localised phase, yet has a covariance (Eq.\ \eqref{eq:tfi-cr}) with a quadratic component in $r/N$.
We believe the likely resolution to this is that in the numerical studies of \cite{ghosh2019manybody}, the total 
disorder strength was not scanned sufficiently (it was in fact held constant).


\subsection{{\textit{p}}-spin models}
\label{subsection:pspinmods}

The family of $p$-spin models has a long history in the field of spin-glasses, both classical~\cite{sherrington1975solvable,derrida1980random,gross1984simplest,gardner1985spin,kirkpatrick1987pspin,wolynes1987pspin} as well as quantum~\cite{goldschmidt1990solvable,nieuwenhuizen1998quantum}. More recently, they have attracted attention in the context of ergodicity breaking and many-body localisation~\cite{baldwin2017clustering,burin2017localization,mukherjee2018many}.
In the following we show that for the  incarnation of the $p$-spin model considered, the 
distributions of Fock-space site energies are multivariate Gaussians, such that the rescaled covariances
$\rho_r\to1$ for finite $r$ in the thermodynamic limit. The criterion identified in
Sec.\ \ref{subsubsec:criterion} then implies that a localised phase, and hence a many-body localisation transition,
should exist. This is subsequently studied and confirmed by exact diagonalisation 
(Sec.\ \ref{subsection:EDpspinmods}).

We write the Hamiltonian for the $p$-spin model as 
\begin{equation}
H^{p\text{-spin}} = H_\text{diag}^{p\text{-spin}}+H_1^{\pd},
\end{equation}
where 
\begin{equation}
H^{p\text{-spin}}_\text{diag} = \frac{1}{N^{\frac{p-1}{2}}}\sum^{\pd}_{\ell_1,\ell_2\cdots,\ell_p}J^{\pd}_{\ell_1,\ell_2\cdots,\ell_p}\sigma^z_{\ell_1}\sigma^z_{\ell_2}\cdots\sigma^z_{\ell_p},
\end{equation}
and $H_1 =\Gamma \sum_{\ell}\sigma_{\ell}^{x}$  (Eq.~\eqref{eq:ham-offdiag}).
The $J_{\ell_1,\ell_2\cdots,\ell_p}$ are independent random variables drawn from a normal distribution with zero mean and standard deviation $W_J$, i.e.\
\begin{equation}
\braket{J_{i_1,i_2,\cdots,i_p}^n}=\begin{cases} 0\,\,\,\,\,\,\,\,\,\,\,\,\,\,\,\,\,\,\,\,\,\,\,\,\,\,\,\,\,\,\,\,\,\,: n\text{ odd}\\ W_J^n\left(\frac{n}{2}-1\right)!!\,\,: n\text{ even}\end{cases}.
\label{eq:moments-J-pspin}
\end{equation}
Note that the random couplings are independent across different permutations of the same set of $p$ spins, meaning
\begin{equation}
\braket{J^{\pd}_{\ell_1,\ell_2\cdots,\ell_p}J^{\pd}_{m_1,m_2\cdots,m_p}} = W_J^2\prod_{u=1}^p\delta_{\ell_u m_u}^{\pd}:=W_J^2\delta_{\bm{\ell m}}^{\pd}
\label{eq:two-point-J-spin}
\end{equation}
(with $\bm{\ell}$  shorthand for the sequence $\ell_1,\ell_2,\cdots,\ell_p$).
We consider for simplicity the case where there is no constant contribution to the Fock-space site energies, i.e.\  $\E_I=\E_I^0+\E_I^\prime$ with $\E_I^0=0$ and 
\begin{equation}
\mathcal{E}_I^\prime=\frac{1}{N^{\frac{p-1}{2}}}\sum^{\pd}_{\ell_1,\ell_2\cdots,\ell_p}J^{\pd}_{\ell_1,\ell_2\cdots,\ell_p}s^{(I)}_{\ell_1}s^{(I)}_{\ell_2}\cdots s^{(I)}_{\ell_p}.
\label{eq:decompose-fs-energy-pspin}
\end{equation}


\subsubsection{{Distribution and correlations of Fock-space site energies}
\label{subsection:EDpspinmods}}

The first step is to show that the $N_\mathcal{H}$-dimensional distribution of $\{\mathcal{E}_I^\prime\}$ over disorder realisations, $P_{N_\mathcal{H}}^d$, is a multivariate Gaussian, whence so too is its $\mathcal{D}$-dimensional marginal distribution. For the distributions to be multivarite Gaussians, Isserlis's theorem~\cite{Isserlis1918} 
(or equivalently Wick's theorem) requires that, for even $n=2q$, arbitrary $n$-point correlations of the Fock-space site energies should reduce to a sum of all possible pairwise two-point correlation functions; i.e.\
\begin{equation}
\braket{\mathcal{E}^\prime_{I_1}\cdots\mathcal{E}^\prime_{I_{2q}}}
=
\sum_{P=1}^{(2q-1)!!}\underbrace{\braket{\mathcal{E}^\prime_{I_{P_1}}\mathcal{E}^\prime_{I_{P_2}}}\cdots\braket{\mathcal{E}^\prime_{I_{P_{2q-1}}}\mathcal{E}^\prime_{I_{P_{2q}}}}}_{\text{product of $q$ two-point terms}}
\label{eq:wick}
\end{equation}
where $P$ labels the different ways of grouping the $2q$ energies into $q$ pairs. For odd $n$, the correlation vanishes trivially.

While Eq.~\eqref{eq:wick} can be proven in general by induction, for brevity we discuss explicitly the case of $q=2$. This corresponds to studying the four-point correlation, which for the distribution of $\{\mathcal{E}_I^\prime\}$ to be a multivariate Gaussian should reduce to 
\begin{align}
\braket{\mathcal{E}^\prime_{I_1}\mathcal{E}^\prime_{I_2}\mathcal{E}^\prime_{I_3}\mathcal{E}^\prime_{I_4}}=&
\braket{\mathcal{E}^\prime_{I_1}\mathcal{E}^\prime_{I_2}}\braket{\mathcal{E}^\prime_{I_3}\mathcal{E}^\prime_{I_4}}+\braket{\mathcal{E}^\prime_{I_1}\mathcal{E}^\prime_{I_3}}\braket{\mathcal{E}^\prime_{I_2}\mathcal{E}^\prime_{I_4}}\nonumber\\
&+\braket{\mathcal{E}^\prime_{I_1}\mathcal{E}^\prime_{I_4}}\braket{\mathcal{E}^\prime_{I_2}\mathcal{E}^\prime_{I_3}}.
\label{eq:wick-4}
\end{align}
To show that Eq.~\eqref{eq:wick-4} holds, let us first define
\begin{equation}
\mathcal{I}_4(\bm{i},\bm{j},\bm{\ell},\bm{m}):=\braket{J_{i_1,\cdots,i_p}J_{j_1,\cdots,j_p}J_{\ell_1,\cdots,\ell_p}J_{m_1,\cdots,m_p}}
\end{equation}
which, using Eqs.~\eqref{eq:moments-J-pspin} and \eqref{eq:two-point-J-spin}, can be simplified to
\begin{equation}
\mathcal{I}_4(\bm{i},\bm{j},\bm{\ell},\bm{m}) = W_J^4(\delta_{\bm{ij}}\delta_{\bm{\ell m}}+\delta_{\bm{i\ell}}\delta_{\bm{j m}}+\delta_{\bm{im}}\delta_{\bm{j\ell}}).
\label{eq:I4-pspin}
\end{equation}
The four-point correlation function, $\braket{\mathcal{E}^\prime_{I_1}\mathcal{E}^\prime_{I_2}\mathcal{E}^\prime_{I_3}\mathcal{E}^\prime_{I_4}}$ can be expressed in terms of $\mathcal{I}_4$ as 
\begin{widetext}
\begin{equation}
\begin{aligned}
\braket{\mathcal{E}^\prime_{I_1}\mathcal{E}^\prime_{I_2}\mathcal{E}^\prime_{I_3}\mathcal{E}^\prime_{I_4}}
&=
\frac{1}{N^{2(p-1)}}
\sum_{\bm{i},\bm{j},\bm{\ell},\bm{m}}
\mathcal{I}_4(\bm{i},\bm{j},\bm{\ell},\bm{m})
[s^{(I_1)}_{i_1}\cdots s^{(I_1)}_{i_p}]
[s^{(I_2)}_{j_1}\cdots s^{(I_2)}_{j_p}]
[s^{(I_3)}_{\ell_1}\cdots s^{(I_3)}_{\ell_p}]
[s^{(I_4)}_{m_1}\cdots s^{(I_4)}_{m_p}]\\
&=
\frac{W_J^4}{N^{2(p-1)}}\sum_{\bm{i},\bm{\ell}}
\left[
\left(\prod_{u=1}^p s_{i_u}^{(I_1)}s_{i_u}^{(I_2)}\right)\left(\prod_{v=1}^p s_{\ell_v}^{(I_3)}s_{\ell_v}^{(I_4)}\right)
+
\left(\prod_{u=1}^p s_{i_u}^{(I_1)}s_{i_u}^{(I_3)}\right)\left(\prod_{v=1}^p s_{\ell_v}^{(I_2)}s_{\ell_v}^{(I_4)}\right)\right.\\
&~~~~~~~~~~~~~~~~~~~~~~~~~~~~~~~~~~~~~~~~~~~~~~~~~~+
\left.\left(\prod_{u=1}^p s_{i_u}^{(I_1)}s_{i_u}^{(I_4)}\right)\left(\prod_{v=1}^p s_{\ell_v}^{(I_2)}s_{\ell_v}^{(I_3)}\right)
\right]
\end{aligned}
\label{eq:four-point-expand}
\end{equation}
\end{widetext}
(using Eq.~\eqref{eq:I4-pspin} in the second line). However, noting that the two-point correlation 
$\braket{\mathcal{E}^\prime_{I_1}\mathcal{E}^\prime_{I_2}}$ can be expressed as
\begin{equation}
\braket{\mathcal{E}^\prime_{I_1}\mathcal{E}^\prime_{I_2}} = \frac{W_J^2}{N^{p-1}}\sum_{\bm{\ell}}\prod_{u=1}^p s_{\ell_u}^{(I_1)}s_{\ell_u}^{(I_2)},
\label{eq:two-point-expand}
\end{equation}
one immediately identifies the four-point correlation in Eq.~\eqref{eq:four-point-expand} as reducing to 
Eq.~\eqref{eq:wick-4},  as required. As mentioned above, such an analysis can be generalised to arbitrary even $n$, and the general result proven by induction. Before providing numerical evidence for $\mathcal{P}_{N_\mathcal{H}}^d$ being a Gaussian, notice that Eq.~\eqref{eq:two-point-expand} is  precisely the correlation function entering the covariance matrix of the  multivariate Gaussian. We thus sketch its derivation.

Each term in the sum in Eq.~\eqref{eq:two-point-expand} is a product over $p$ spins, not necessarily distinct.
Let the Hamming distance between Fock-space sites $I_1$ and $I_2$ be $r$ such that, out of the $p$ spins, $x$ of them are chosen from the $r$ spins which differ in $I_1$ and $I_2$, and the remaining $p-x$ are chosen from the $N-r$ spins which are the same. There are $r^x$ and $(N-r)^{p-x}$ ways  respectively of chosing the former and latter (as repeated indices are allowed), and they contribute $(-1)^x$ and $(+1)^{p-x}=1$ to Eq.\ \eqref{eq:two-point-expand}.
And one can clearly choose the $x$ spins out of $p$ in $\binom{p}{x}$ ways. Taking all this into consideration, the two-point correlation can thus be written as
\begin{equation}
\begin{aligned}
\braket{\mathcal{E}^\prime_{I_1}\mathcal{E}^\prime_{I_2}} &= \frac{W_J^2}{N^{p-1}}\sum_{x=0}^p\binom{p}{x} (N-r)^{p-x}(-r)^x\\
&=W_J^2 N\left(1-2\frac{r}{N}\right)^p.
\end{aligned}
\label{eq:two-point-cor}
\end{equation}
It is important to note that  this covariance depends only on the Hamming distance between the two states $I_1$ and $I_2$; no other specifics of the state are required. One thus concludes that for the $p$-spin model, 
$C_d(r)=W_J^2 N\left(1-2r/N\right)^p$.

\begin{figure}[b]
\includegraphics[width=\columnwidth]{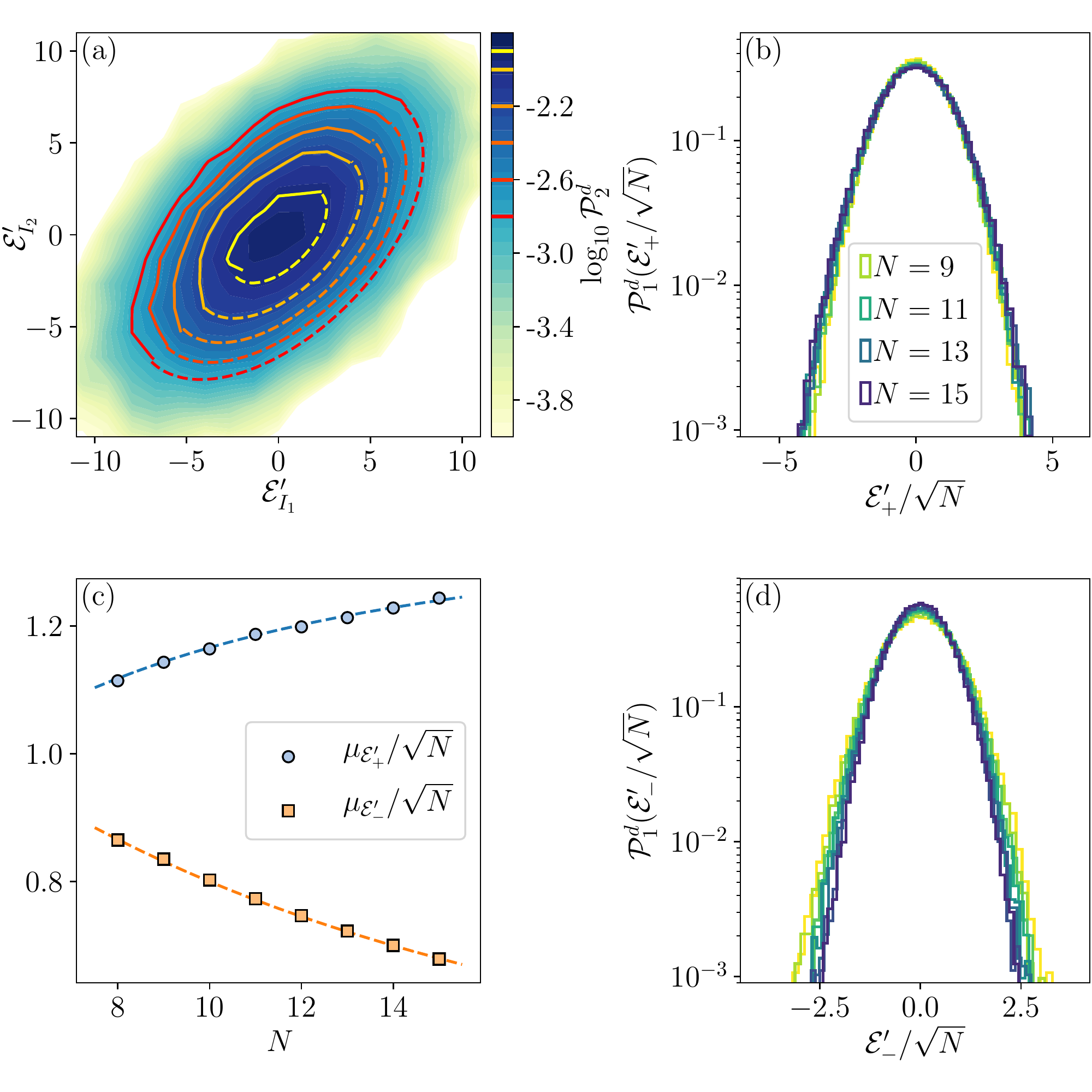}
\caption{Results for the distribution of Fock-space site energies over disorder realisations, for a 
$p=2$-spin model with $W_{J}=1$, using a bivariate case $\mathcal{P}_2^d$ with $\ket{I_1}=\ket{\up\cdots\up\up\up}$ and 
$\ket{I_2}=\ket{\up\cdots\up\dn\dn}$ ($r=2$). (a) The distribution is shown as a contour plot, with elliptical contours 
indicating its Gaussian nature. This is confirmed by comparing the highlighted contours, in which the solid part corresponds to the numerical distribution and the dashed part to a bivariate Gaussian with the same covariance. In (b) and (d), the univariate distributions of $\E_\pm^\prime/\sqrt{N}$ for different system sizes, $N$,  also show Gaussian behaviour, as indicated by their inverted parabolic shapes on log-linear axes. The standard deviations of these distributions,  $\mu_{\E_\pm^\prime}/\sqrt{N}$, are shown in (d), where the dashed lines correspond to the analytic prediction  $\mu_{\E_\pm^\prime}^{2}=C_d(0)\pm C_d(r)$, with $C_d(r)$ given by  Eq.~\eqref{eq:two-point-cor}. 
Results were obtained  with $5\times 10^4$ disorder realisations for all $N$; and for (a), $N=15$. }
\label{fig:pd-pspin}
\end{figure}

\begin{figure}
\includegraphics[width=\columnwidth]{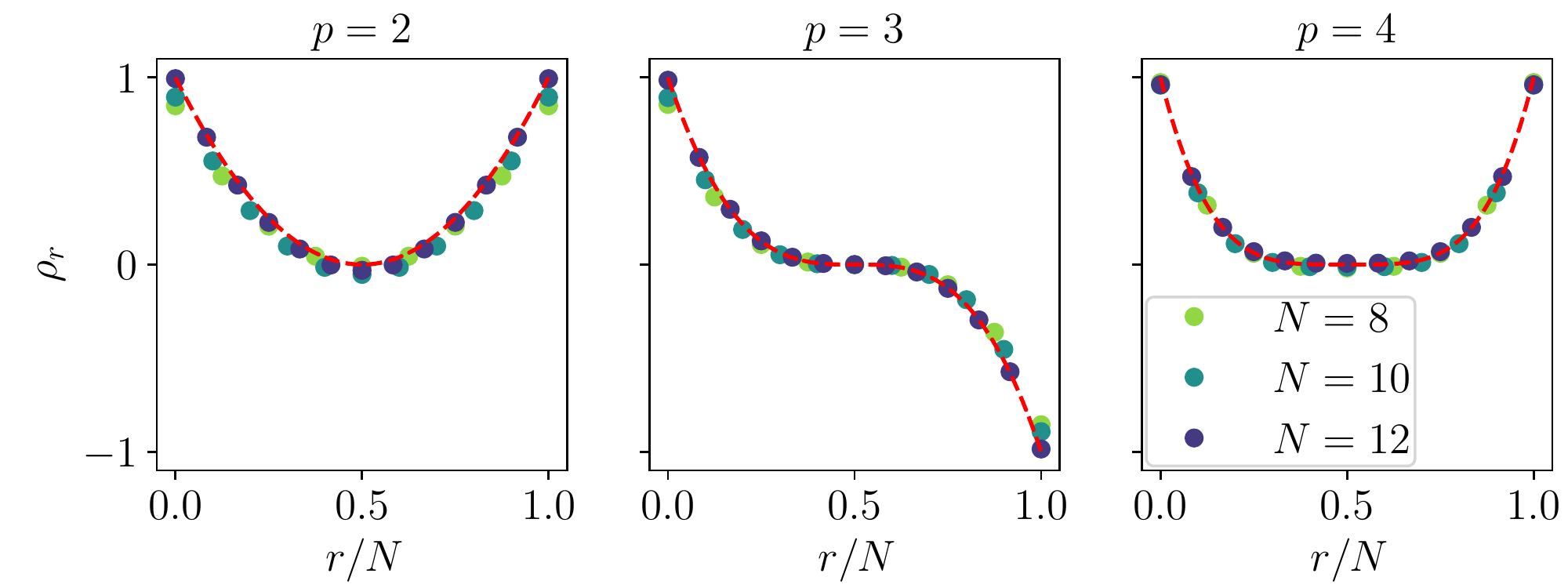}
\caption{For $p$-spin models, the numerically calculated rescaled covariance $\rho_r=C(r)/C(0)$ is shown 
as a function of $r/N$, for different values of $N$ and for $p=2-4$. The red dashed line in each case is the analytical prediction $\rho_r = (1-2r/N)^p$ from Eq.\ \eqref{eq:two-point-coragain}.}
\label{fig:correl-p-spin}
\end{figure}

We turn now to numerical study of the distributions, focussing again on $\mathcal{P}_2^d$, 
in parallel to the analysis and discussion in Sec.\ \ref{subsubsection:NN-TFI-dists} for the short-ranged model.
The results are shown in Fig.~\ref{fig:pd-pspin}.  The numerical $\mathcal{P}_2^d(\E^\prime_{I_1},\E^\prime_{I_2})$ shows
very good agreement with a bivariate Gaussian with the elements of its covariance matrix given 
by Eq.~\eqref{eq:two-point-cor}. This is also evident in the distributions $\mathcal{P}_1^d(\E^\prime_\pm)$ (with
$\E_\pm^\prime =(\E_{I_{1}}^\prime\pm\E_{I_{2}}^\prime)/\sqrt{2}$ as before),
which are clear univariate  Gaussians, as shown  in panels (b) and (d) of Fig.~\ref{fig:pd-pspin}.
Their standard deviations  $\mu_{\E^\prime_\pm}/\sqrt{N}$, shown in panel (d), are moreover in excellent 
agreement with the analytically derived form of  $\mu_{\E^\prime_\pm}^{2}=C(0)\pm C(r)$ (with
$C(r)$ from Eq.~\eqref{eq:two-point-cor}). Overall, the numerical results both confirm that $\mathcal{P}^d_2$ is a bivariate Gaussian, and that the elements of the covariance matrix depend only on the Hamming distance between the corresponding Fock-space sites, as given by Eq.~\eqref{eq:two-point-cor}.

Since there is no disorder-independent contribution to the Fock-space site energies, the  distribution over both disorder realisations and Fock space, $\P_\mathcal{D}\equiv\P_\mathcal{D}^d$, is thus a Gaussian with the same covariance matrix
\begin{equation}
C(r)=C_d(r)=W_J^2N\left(1-2\frac{r}{N}\right)^p,
\label{eq:two-point-coragain}
\end{equation}
from which $\rho_r=C(r)/C(0)=(1-2r/N)^p$. This polynomial behaviour in $r/N$ is well captured numerically, as seen from the results for $\rho_{r}$ \emph{vs} $r/N$ shown in Fig.~\ref{fig:correl-p-spin} for the cases $p=2,3,4$.

The particular significance of this result is that $\rho_r\to 1$ for finite values of
$r$ in the thermodynamic limit $N\to\infty$. For any finite $p$, the $p$-spin models thus fall into the class of models with maximally correlated Fock-space site energies. From the analysis in Sec.~\ref{subsec:maximally}, we thus predict them to posses a many-body localised phase and an ensuing many-body localisation transition.


\subsubsection{Results from exact diagonalisation}
\label{subsubsection:ED-pspinmods}

\begin{figure*}
\includegraphics[width=\linewidth]{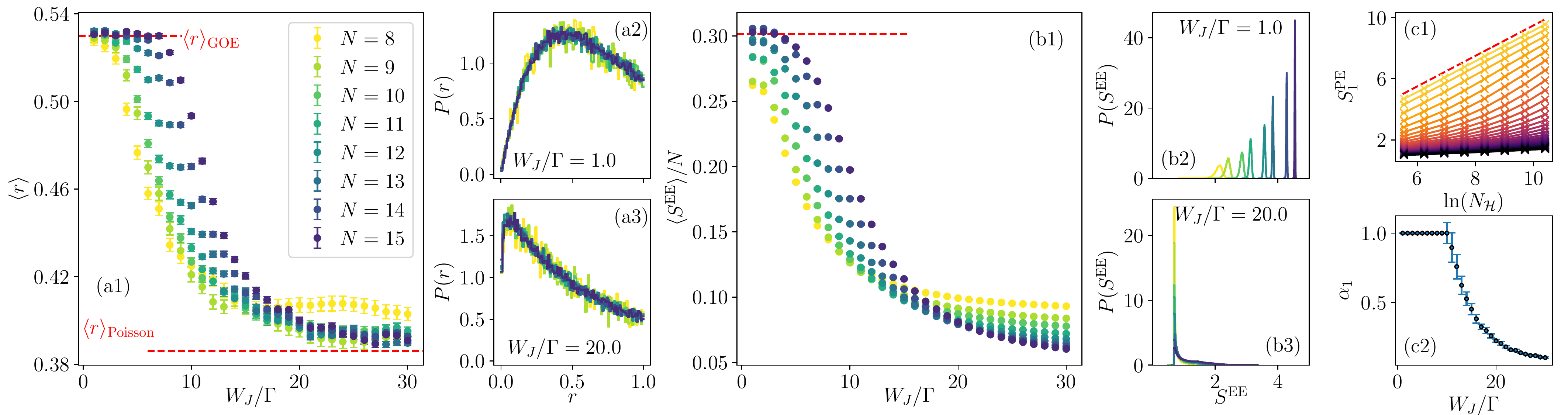}
\caption{Exact diagonalisation results for a $p=2$-spin model. Panels show the same diagnostics as Fig.~\ref{fig:ed-nn}, to which the results are qualitatively similar. The mean level-spacing ratio $\braket{r}$ \emph{vs} $W_J/\Gamma$ (panel 
(a1)) goes from the the GOE value to the Poisson value on increasing $W_J/\Gamma$; with crossing {}of the data for various $N$ indicating the presence of a transition. Panels (a2) and (a3) show clear Wigner-Dyson and Poisson distributions of $r$ in the delocalised and localised phases respectively. The half-chain entanglement entropy, 
$\braket{S^\mathrm{EE}}$ (panel (b1)), also goes from a volume law to an area law on increasing $W_J/\Gamma$, as further exemplified in their distributions, (b2) and (b3). The first participation entropy, $S^\mathrm{PE}_1$ \emph{vs} 
$\ln(N_\mathcal{H})$, is shown in (c1) for various disorder strengths (lighter colours denote weaker disorder strengths).
The exponent $\alpha_1$, defined in Eq.~\eqref{eq:PEscaling}, deviates from unity only above a threshold disorder strength ((c2)), indicating a transition to the localised phase. Data were obtained by averaging over 1000 disorder realisations, with statistical errors estimated using a standard bootstrap error analysis.}
\label{fig:ed-pspin}
\end{figure*}

Results from exact diagonalisation are shown in Fig.~\ref{fig:ed-pspin}, for the representative case of $p=2$.
These are qualitatively similar to those for the disordered short-ranged spin chain (Fig.~\ref{fig:ed-nn}), 
and again show the hallmarks of a stable many-body localised phase and a localisation transition. 

The mean level-spacing ratio, $\braket{r}$, crosses over from $\braket{r}_\text{GOE}$ to $\braket{r}_\text{Poisson}$ 
as $W_J/\Gamma$ is increased, with the data for various $N$ exhibiting a  crossing, see panel (a1). 
For values of $W_J/\Gamma$ representative of delocalised and localised phases respectively, the distributions $P(r)$ of the level-spacing ratio clearly follow that of the Wigner-Dyson and Poisson distributions, see panels (a2) and (a3). 
Similarly, the half-chain entanglement entropy scaled by system size, $\braket{S^\mathrm{EE}}/N$, approaches the known 
value for an ergodic state at low $W_J/\Gamma$ with increasing $N$, but crosses over to area-law behaviour at high 
$W_J/\Gamma$, with the data for various $N$ again showing a crossing, see panel (b1). 
The distributions $P(S^\text{EE})$ of the entanglement entropy, shown in panels (b2) and (b3), also exemplify this 
behaviour. Finally, the first participation entropy $S^{\text{PE}}_1$ again scales linearly with $\ln(N_\mathcal{H})$, with the slope $\alpha_{1}$ (Eq.\ \eqref{eq:PEscaling}) sticking to unity for a range of  $W_J/\Gamma$, before decreasing below it as the transition is crossed. Note that close to the transition the $S^{\text{PE}}_1$ \emph{vs} 
$\ln(N_\mathcal{H})$ curves bend somewhat, reflecting the amplified finite-size effects expected in such infinite-range models (as seen also in the greater drift in crossing points compared to the short-ranged
spin model).

As found for the disordered quantum Ising chain (Sec.\ \ref{subsection:NNTFI}), the numerical diagnostics for
the $p=2$-spin model thus concurrently predict a localised phase and a localisation transition, in agreement with the theory presented in Sec.~\ref{subsec:maximally}.

Before concluding our discussion of $p$-spin models, we make two remarks about the generality of the results. First, in 
demonstrating Isserlis's theorem for the Fock-space site energies, we exploited the fact that the independent random couplings $J_{\bm{\ell}}$ were normally distributed.  The result is not however confined to that case. It can be shown that for any distribution of $J_{\bm{\ell}}$ which has well-defined moments, the corrections to Isserlis's theorem 
 vanish in the thermodynamic limit. 

Second, we considered explicitly the case where the couplings $J_{\bm{\ell}}$ had a vanishing mean 
(Eq.\ \eqref{eq:moments-J-pspin}). In these infinite-ranged $p$-spin models, a non-zero mean in fact leads to a non-Gaussian form for the distribution $\P_{\mathcal{D}}^{F}$ of the $\E_{I}^{0}$ over the Fock space (while 
$\P_{\mathcal{D}}^{d}$ remains Gaussian); as discussed further in Appendix \ref{app:nongaussianpspin}.
The model nevertheless remains in the  maximally correlated class, because the site energies of neighbouring Fock-space sites still differ by an $\mathcal{O}(1)$ number. One thus concludes that the model still supports a localised phase and hosts a localisation transition (for further details, see also  Appendix~\ref{app:nongaussian}).


\subsection{REM with exponentially decaying correlations}
\label{subsection:expREM}

\begin{figure*}
\includegraphics[width=\linewidth]{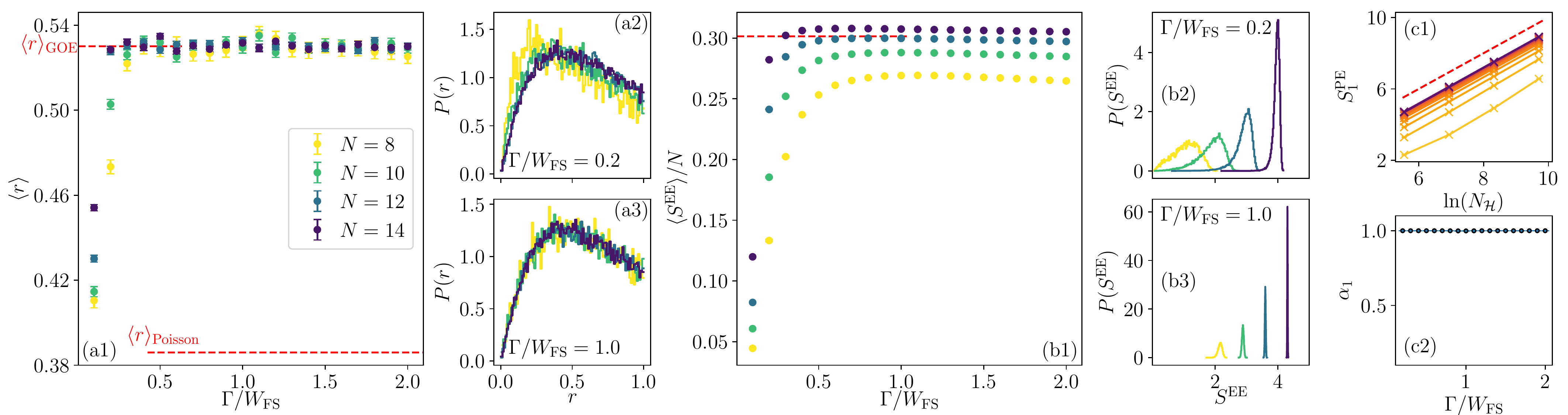}
\caption{Exact diagonalisation results for the ExpREM model (with $\xi=1$). Panels show the same quantities as in Figs.~\ref{fig:ed-nn} and \ref{fig:ed-pspin}, but with qualitatively different results. Signifying  an ergodic phase,
the mean level-spacing ratio $\braket{r}$ \emph{vs} $\Gamma/W_{\tot}$, panel (a1), and the half-chain entanglement entropy 
$\braket{S^\mathrm{EE}}/N$ \emph{vs} $\Gamma/W_{\tot}$ ((b1)), persist respectively at their GOE and volume-law values at all $\Gamma/W_\tot$. The distributions of $r$ and $S^\mathrm{EE}$ (panels (a2)-(a3) and (b2)-(b3)) are likewise qualitatively similar at both high and low $\Gamma/W_\tot$,  exhibiting respectively the Wigner-Dyson and 
volume-law behaviours expected for an ergodic system.  Panel (c1) shows the first participation entropy, 
$S^\mathrm{PE}_1$  \emph{vs} $\ln(N_\mathcal{H})$,  for various disorder strengths (lighter colours denote weaker disorder strengths, and the dashed line shown has unit gradient).
It grows linearly with $\ln \nh$, with a slope $\alpha_1=1$ for all values of $\Gamma/W_\tot$
(panel (c2)),  indicating a delocalised phase exclusively.   Data were obtained by averaging over 1000 disorder realisations, with statistical errors estimated using a standard bootstrap error analysis.}
\label{fig:ed-exprem}
\end{figure*}

In a model with local interactions, the Hamiltonian can only possess a number of independent random variables
that scales polynomially with $N$. It is thus inevitable that the associated Fock-space site energies 
are strongly correlated, since the number of them is exponentially large in $N$; as indeed is the case for both the
disordered quantum Ising chain and the $p$-spin models.
It is reasonable to conjecture that a model which does not have maximally correlated Fock-space site energies 
cannot be described by a local Hamiltonian. Hence, to test numerically the absence of localisation in such a model 
-- one in which $\rho_{r}$ for finite $r$ does not tend to unity in the thermodynamic limit -- we
construct such a model on the Fock space directly.

The off-diagonal part of the Hamiltonian, $H_1$, remains given by Eq.~\eqref{eq:ham-offdiag}. The Fock-space site energies are given by a set of random numbers drawn from a $N_\mathcal{H}$-dimensional Gaussian distribution, with a covariance matrix whose elements again depend only on the Hamming distance between the Fock-space sites, as
given by
\begin{equation}
C(r)=W_{\tot}^2 N \exp[-r/\xi]
\label{eq:cr-exprem}
\end{equation}
with $\xi$ independent of $N$. Due to the exponentially decaying correlations, we refer to this model as the ExpREM.
Its Hamiltonian can then be written as
\begin{equation}
H_\text{ExpREM}^{\pd} = \sum_{I}\mathcal{E}_I\ket{I}\bra{I}+\Gamma\sum_\ell\sigma^x_\ell ,
\label{eq:ham-exprem}
\end{equation}
where the distribution of Fock-space site energies is
\begin{equation}
\P_{N_\mathcal{H}}^{\pd}(\bm{\mathcal{E}}) = \frac{1}{\sqrt{(2\pi)^{N_\mathcal{H}}\vert\mathbf{C}\vert}}\exp\left[-\frac{1}{2}\bm{\mathcal{E}}^{\mathrm{T}} \cdot \mathbf{C}^{-1}\cdot \bm{\mathcal{E}}\right]
\end{equation}
with $C(r)$ given by Eq.~\eqref{eq:cr-exprem}. Appendix~\ref{app:exprem} outlines the numerical construction of 
the model.  Note that $\rho_r=C(r)/C(0)=\exp[-r/\xi]\nrightarrow 1$ in the thermodynamic limit, whence the results of Sec.~\ref{subsubsec:arbitcor} predict that this model does not possess a localised phase.

Exact diagonalisation results for the ExpREM are shown in Fig.~\ref{fig:ed-exprem}. They 
are qualitatively different from those obtained for the disordered Ising chain and $p$-spin models. 
There is no $N$-dependent crossing of the data for the mean level-spacing ratio or the bipartite entanglement entropy. 
$\braket{r}$ stays at its GOE value for almost all values of $\Gamma/W_\tot$, and even for the smallest 
value of such used in the numerical calculations it shows a systematic drift with increasing $N$ towards the GOE value, see panel (a1). This is further exemplified by the distributions of $P(r)$ showing Wigner-Dyson behaviour for both high and low values of $\Gamma/W_\tot$.  $\braket{S^\mathrm{EE}}/N$,  shown in  panel (b1), likewise shows a 
systematic drift with increasing $N$ towards the ergodic volume-law value. Finally, the first participation entropy 
$S_1^\mathrm{PE}$ shows a linear behaviour with $\ln(N_\mathcal{H})$, with a slope $\alpha_{1}=1$  for all 
$\Gamma/W_\tot$ (panel (c1)), indicating that the model is always delocalised.

The numerical results are thus consistent with the absence of a localised phase  of the ExpREM model, in agreement with  the predictions of the self-consistent theory based on the behaviour of $\rho_r$.


\section{Discussion 
\label{sec:discussion}}

A brief summary of the paper is first in order (for more details see the Overview, Sec.~\ref{sec:overview}). Considering the problem of many-body localisation from the natural perspective of Fock space, our central aim has been to understand the minimal conditions on the Fock-space correlations required for a many-body localised phase to be stable. Any many-body Hamiltonian maps onto a tight-binding model on the Fock space graph, with correlated disorder in its diagonal elements (the Fock-space site energies). A first important step was thus to establish the nature of correlations between these site energies, encoded in their joint probability distributions (Sec.~\ref{sec:FSgraph}). 
Together with the fact that the connectivity of a Fock- space site is extensive in $N$, these distributions then formed key ingredients in a probabilistic mean-field theory for localisation on the Fock space (Sec.~\ref{sec:fockspacelocalisation}). This centres on a self-consistent determination of the probability distribution for the imaginary part of the self-energy corresponding to the local Fock-space propagator; with a natural focus on the stability of the many-body localised phase, and band centre states. Expressed in terms of the reduced covariance $\rho_r$, the theory predicts that a stable many-body localised phase arises in what we term the maximally correlated limit, where $\rho_r\to1$ for finite $r$ in the thermodynamic limit $N\to\infty$.
The same theory also predicts that a many-body localised phase cannot be stable in either the wholly uncorrelated limit of $\rho_r =0$ for all non-zero $r$ -- capturing as such the known result for the quantum random energy model 
(QREM)~\cite{laumann2014many,baldwin2016manybody} -- or for general $\rho_r\in(0,1)$, which was shown to belong to the QREM class. Hence our general conclusion: a many-body localised phase is stable only in the class of models that are maximally correlated.
  
The theoretical predictions were then tested (Sec.~\ref{sec:microscopic}) by detailed numerical study of three classes of microscopic models, using exact diagonalisation. In each case, these results were found to concur with the theory.

Two further points of discussion are germane. First, the fact that a disordered many-body Hamiltonian maps
onto a tight-binding problem on a disordered graph (Fock-space)  with an extensive  connectivity ($\propto N$), should not obscure the fact that it is the correlations on the graph that render the problem fundamentally different from that of conventional Anderson localisation on high-dimensional graphs/lattices. In that regard, it is the `extreme' QREM limit which is the  essential many-body analogue of the Anderson localisation problem in high dimensions; for in both cases the associated site energies -- in Fock-space  on the one hand and real-space on the other -- 
are completely uncorrelated, independent random variables. It is for this common underlying reason that the QREM does not host a many-body localised phase in the thermodynamic limit, and that the critical disorder for the Anderson localisation problem scales to infinity as the connectivity of the lattice diverges~\cite{abou-chacra1973self}, signalling 
the absence of an Anderson localised phase in that limit.

Relatedly, it is also interesting to note that within the context of Anderson localisation, suitably engineered correlations between the one-body site energies can induce localisation~\cite{nosov2018correlation}. In fact, 
Ref.~\cite{nosov2018correlation} discusses the presence of non-ergodic single-particle states whose spatial
support scales in a non-trivial fashion with the system size; which is redolent of the structure of many-body localised eigenstates on the Fock space.  Understanding the connections, should they exist, between correlations in the Fock-space site energies and the scaling of the Fock-space support of localised eigenstates, is an interesting question for future consideration.

In recent work focussing on short-ranged models, a classical percolation model on the Fock space was constructed, acting as a proxy for quantum many-body localisation~\cite{roy2018exact,roy2018percolation}. The percolation picture was based on a competition between Fock-space site-energy differences, and the hopping energy scale on the Fock-space. 
It is thus  natural to ask how the effect of correlations, or their absence, is manifest in the percolation picture. 
While a detailed analysis of this is a subject for future research, a heuristic picture can be given as follows. 
Corresponding to a particular spin-flip, there are exponentially many (in $N$) links on the Fock space. In short-ranged models, however, there are only an $\mathcal{O}(1)$ number of energy scales associated with the spin flip,  arising from the configurations of the $\mathcal{O}(1)$ neighbours with which the spin in question interacts. As a result, even a few of these energy scales becoming non-resonant is enough to suppress an exponentially large number of hoppings on the Fock space. This was shown to freeze blocks of spins collectively, and is the mechanism underlying the transition to the localised/non-percolating phase. In the uncorrelated (QREM-like) case by contrast, each of the exponentially large number of Fock-space links corresponding to a spin-flip is independent. Hence, not only is there no obvious collective effect but, since there are an exponentially large number of independent ways in which a spin can flip, the probability of a spin (or block of spins) freezing is vanishingly small in the thermodynamic limit,  which thus precludes localisation.

As a concluding remark, we reiterate the core message of this work, namely that the presence of strong  correlations on the Fock-space is intimately connected to the origins of many-body localisation. In a forthcoming work, it will also be shown that strong ergodicity breaking in Fock space can arise via a complementary mechanism in a different class of models, namely those with imposed kinetic constraints~\cite{roy2019strong}.

\begin{acknowledgments}
We are grateful for helpful discussions with J. T. Chalker, I. Creed, A. Lazarides, A. Nahum and S. Welsh. 
DEL expresses his warm thanks to members of the Physics Department of the Indian Institute of Science, Bangalore,
for their kind hospitality, and to the Infosys Foundation for support.
This work was supported by EPSRC Grants No. EP/N01930X/1 and EP/S020527/1.
\end{acknowledgments}


\appendix
\section{Non-normally distributed, maximally correlated Fock-space site energies \label{app:nongaussian}}

Here we consider again the self-consistent theory in the maximally correlated limit, but for a case where the distribution of the Fock-space site energies is not a Gaussian. We start with the expression from Eq.~\eqref{eq:Fk-strong} (with $\mu_{\mathcal{E}}^{2} =W_{\tot}^{2}N$, Eq.\ \eqref{eq:WFSdef}),
\begin{equation}
F_k^{\pd}(k)=\int dx_{I_0}^{\pd}\P_1^{\pd}(x_{I_0}^{\pd})\exp\left(\frac{ik\kappa}{W_{\tot}^{2} x_{I_0}^2}\right).
\label{eq:A1}
\end{equation}
Note however that $F_k(k)$ is a function solely of $\tilde{k}=k\lambda$, $F_k(k):=\hat{f}(k\lambda)$, where we define 
\begin{equation}
\lambda=\frac{\kappa}{W_{\tot}^{2}} = \frac{\Gamma^2}{W_{\tot}^2}(1+y_\typ).
\end{equation}
Since $F_y(y,y_\typ)$ is a Fourier transform of $\hat{f}(k\lambda)$,
\begin{equation}
F_y^{\pd}(y,y_\typ) = \frac{1}{2\pi\lambda}\int d\tilde{k}\,e^{-i\tilde{k}y/\lambda}~\hat{f}(\tilde{k}),
\end{equation}
it thus has the scaling form $F_y(y,y_\typ)=\lambda^{-1}f(y/\lambda)$, with $f$ the inverse 
transform of $\hat{f}$ and $\int_0^\infty dy f(y)=1$ due to normalisation. The self-consistency condition, 
$\ln y_\typ=\int_{0}^{\infty} dy~ F_y(y,y_\typ)\ln y$, thus takes the form 
\begin{equation}
\begin{aligned}
\ln y_{\typ}^{\pd}=&\int_0^\infty dy\, f(y)\ln(y\lambda)=\ln \lambda + \underbrace{\int_0^\infty dy f(y)\ln y}_{c}\\
\Rightarrow\, &y_\typ^{\pd}=\frac{\Gamma^2 e^c}{W_{\tot}^2-\Gamma^2 e^c}.
\end{aligned}
\end{equation}
Note that there is no $N$-dependence in the above equation. In direct parallel to the discussion following
Eq.\ \eqref{eq:ytyp-strong}, it follows that a transition thus exists provided $c$ is finite, and a localised phase is stable for $W_\tot>\Gamma e^{c/2}$.

A subsequent natural question arises: what can be deduced about the behaviour of $F_y(y,y_\typ)$ without making any strong assumptions about the form of $\P_1$? Note that the Fock-space site energies are always referred relative to their mean, and hence $\P_1(0)\neq 0$. This is the only ingredient we use, and it is sufficient to show that $F_y$ has the L\'evy tail, $\propto y^{-3/2}$, characteristic of the localised phase~\cite{logan2019many,roy2019self}.
To see this, let us recast $F_k$ (Eq.\ \eqref{eq:A1}) as 
\begin{equation}
F_k^{\pd}(k) = 1+\int dx\, \P_1^{\pd}(x)\left[\exp\left(ik\lambda/x^2\right)-1\right].
\end{equation}
Defining $u=x/\sqrt{\lambda\vert k\vert}$, the leading low-$k$ behaviour of $F_k$ (as we are interested in the tails of the distribution $F_y$) is given by
\begin{equation}
F_k^{\pd}(k)\overset{k\to 0}{\sim} 1- [1-i\mathrm{sgn}(k)]\sqrt{2\pi\P_1^2(0)\frac{\kappa}{W_{\tot}^2}\vert k\vert}.
\end{equation}
The fact that the leading low-$k$ behaviour of $[F_k(k)-1]\propto \vert k\vert^{1/2}$ 
guarantees that the large-$y$ behaviour of $F_y(y,y_\typ)$ is $\propto y^{-3/2}$, whence the L\'evy tail. Note that this also means that $c=\int_0^\infty dy~ f(y)\ln y$ is indeed finite, ensuring a well-defined localisation criterion.

\section{Non-Gaussian distributions in \textit{p}-spin models}
\label{app:nongaussianpspin}

For $p$-spins models, we show here that if there exists a constant component to the random couplings,  leading to a non-vanishing $\mathcal{E}_I^0$, then the distribution over the Fock space, $\P^F_\mathcal{D}(\{\E_I^0\})$, is not of normal form. For simplicity, we consider the simplest case of $p=2$ and $\mathcal{D}=1$. 

With $J$ denoting the constant contribution to the exchange couplings, the disorder-independent part of the Fock-space site energy is simply
 \begin{equation}
\E_I^0 = \frac{J}{\sqrt{N}}\sum_{\ell_1,\ell_2}s_{\ell_1}^{(I)}s_{\ell_2}^{(I)}.
\end{equation}
Since the $p$-spin model is by construction an infinite-range model, $\E_I^0$ depends only on the total number of up and down spins in the configuration $I$, and not on their spatial locations.  Denote these numbers by $N_{I}^\up$ and 
$N^\dn_{I}=N-N_{I}^\up$.  Recognising that $\sum_{\ell_{1}}\ssh[I]{\ell_{1}} = (N_{I}^\up -N_{I}^\downarrow)$, 
$\E_I^0$ is given simply by
\begin{equation}
\label{eq:EI0-pspin-J}
\E_I^0 = \frac{J}{\sqrt{N}}\left[N_{I}^{\up}-N^\downarrow_{I}\right]^2
=\frac{J}{\sqrt{N}}\left[N-2N^\up_{I}\right]^2.
\end{equation}
The probability density for a state with $N^\up$ up spins is  $P(N^\up)=2^{-N}\binom{N}{N^\up}$. As the thermodynamic limit is approached, $P(N^\up)$ tends to a Gaussian, $\mathcal{N}(N/2,N/4)$. Transforming from the distribution of 
$N^\up$ to that of $x_I^0=\E_I^0/\sqrt{N}$ using Eq.~\eqref{eq:EI0-pspin-J}, one then arrives at 
\begin{equation}
\P_1^F(x_I^0) = \frac{1}{\sqrt{2\pi J x_I^0}}\exp\left(-\frac{x_I^0}{2J}\right).
\end{equation}
This is of non-Gaussian form
in $x_I^0$, though it decays exponentially, such that all its moments are 
finite. In consequence, the full distribution $\mathcal{P}_1 = \mathcal{P}_1^d\ast\mathcal{P}^F_1$ over both disorder realisations and Fock space is itself non-Gaussian. Nevertheless,  because the model remains in the  maximally correlated class, it still supports a localised phase and hosts a localisation transition, for the reasons explained in
Appendix~\ref{app:nongaussian}.


\section{Construction of the ExpREM model \label{app:exprem}}

Here we sketch the algorithm for the generation of normally distributed Fock-space site energies with arbitrary correlations, which was used to construct the ExpREM model in Sec.\ \ref{subsection:expREM}.
Consider an $N_\mathcal{H}$-dimensional matrix $\mathbf{R}$, such that its $(i,k)^\mathrm{th}$ element $\mathbf{R}_{IK}=r_{IK}$ is the Hamming distance between the Fock-space sites $I$ and $K$. For a desired covariance function of Hamming distance, $C(r)$, the covariance matrix $\mathbf{C}$ can simply be constructed element-wise as
$\mathbf{C}_{IK}=C(\mathbf{R}_{IK})$. 
Let there be a similarity transformation which diagonalises $\mathbf{C}$, 
\begin{equation}
\mathbf{D} = \mathbf{U}^{\mathrm{T}} \cdot\mathbf{C}\cdot \mathbf{U}
\label{eq:diagC}
\end{equation}
where $\mathbf{D}$ is a diagonal matrix with non-negative entries. Now define a $N_\mathcal{H}$-dimensional column vector, $\mathbf{e}$, where each element is independently chosen from a standard normal distribution 
$\braket{e_Ie_K}=\delta_{IK}$. The correlated set of Fock-space site energies is then simply given as 
\begin{equation}
\bm{\E} = \mathbf{W}\cdot\mathbf{e}, \,\,\,\, \mathbf{W}=\mathbf{U}\sqrt{\mathbf{D}}.
\end{equation}
To see this, note that
\begin{equation}
\begin{aligned}
\braket{\E_I^{\pd}\E_K^{\pd}}&=\sum_{I^\prime K^\prime}U_{II^\prime}^{\pd}\sqrt{D_{I^\prime I^\prime}^{\pd}}U_{KK^\prime}^{\pd}\sqrt{D_{K^\prime K^\prime}^{\pd}}\braket{e_{I^\prime}^{\pd}e_{K^\prime}^{\pd}}\\
&=\sum_{I^\prime}U_{I I^\prime}^{\pd}D_{I^\prime I^\prime}^{\pd}U_{KI^\prime}^{\pd}=(\mathbf{U}\cdot \mathbf{D}\cdot\mathbf{U}^\mathrm{T})_{IK}^{\pd}\\
\Rightarrow \braket{\E_I^{\pd}\E_K^{\pd}}&=\mathbf{C}_{IK}^{\pd},
\end{aligned}
\end{equation}
as desired.

\onecolumngrid
\section{Glossary of symbols}
\label{section:glossary}
Table~\ref{tab:symbols} presents a list and description of some of the symbols appearing frequently in the paper.

\begin{table}
\begin{center}
\begin{tabular}{|c|c|}
\hline
Symbol & Description\\
\hline
$\nh=2^N$ & Fock-space dimension for system with $N$ spins-1/2\\
$\E_I$ & Fock-space site energy of site $I$\\
$\E_I^\prime$ & disorder-dependent part of $\E_I$\\
$\E_I^0$ & disorder-independent part of $\E_I$\\
$\P^d_\mathcal{D}$ & $\mathcal{D}$-dimensional distribution over disorder\\
$\mathbf{C}_d$ & covariance matrix corresponding to $\P^d_\mathcal{D}$\\
$\P^F_\mathcal{D}$ & $\mathcal{D}$-dimensional distribution over Fock space\\
$\mathbf{C}_F$ & covariance matrix corresponding to $\P^F_\mathcal{D}$\\
$\P_\mathcal{D}$  & $\P^d_\mathcal{D}\ast\P^F_\mathcal{D}$, distribution over disorder and Fock space\\
$\mathbf{C}$ & covariance matrix corresponding to $\P_\mathcal{D}$\\
$r_{IK}$ & Hamming distance between sites $I$ and $K$\\
$C(r)$ &  covariance as a function of Hamming distance $r$\\
$\mu_\E$ & standard deviation of the distribution of $\E_I$\\
$W_\tot$ &$\mu_\E/\sqrt{N}$,  effective disorder strength on Fock space\\
$\rho_r$ & $C(r)/C(0)$, rescaled covariance\\
$x_I$ & $\E_I/\mu_\E$, rescaled Fock-space site energy\\
$G_I$ & local Fock-space propagator\\
$\Sigma_I$ & local Fock-space self-energy\\
$\Delta_I$ & imaginary part of local Fock-space self-energy\\
$y_I$ & $\Delta_I/\eta$, rescaled $\Delta_I$ with regulator $\eta=0^+$\\
$F_\Delta$ and $F_y$ & distributions of $\Delta_I$ and $y_I$\\
$F_k$& Fourier transform of $F_y$\\
$\gamma$ & Euler-Mascheroni constant, $\gamma = 0.5772\cdots$\\
\hline
\end{tabular}
\end{center}
\caption{List of common symbols and their descriptions.}
\label{tab:symbols}
\end{table}
\twocolumngrid
\bibliography{refs}
\end{document}